\documentclass[prx,twocolumn,groupedaddress]{revtex4-2}

\usepackage{amsmath,amsfonts,amssymb,graphicx,xcolor,ulem}

\newcommand{\ve}[1]{\mathbf{#1}}

\begin{document}



\title{ 
Barrier-crossing  and energy  relaxation
dynamics  of non-Markovian inertial systems
 connected  via analytical Green-Fokker-Planck approach
  }


\author{Roland R. Netz}
\affiliation{Fachbereich Physik, Freie Universit\"at Berlin, 144195 Berlin, Germany}


\date{\today}

\begin{abstract}
From numerical simulations it is known that the barrier-crossing time of a non-Markovian one-dimensional reaction coordinate 
with a single exponentially decaying  memory function exhibits a memory-turnover:
 for intermediate values of the memory decay  time the barrier-crossing time
  is reduced compared to the Markovian limit and for long memory times increases quadratically with the memory time when 
  keeping the total integrated friction and the mass constant.
 The intermediate memory acceleration regime is accurately predicted by Grote-Hynes theory, for the asymptotic  long-memory slow-down behavior
  no systematic  analytically tractable theory is available. 
  Starting from the Green function for a general inertial (i.e. finite-mass)
   non-Markovian Gaussian reaction coordinate in a harmonic well, 
 we derive by an exact mapping a generalized Fokker-Planck equation with a time-dependent effective diffusion constant. 
 To first order in a systematic cumulant expansion 
 we derive an analytical Arrhenius expression for the barrier-crossing time 
 with the pre-exponential factor  given by the energy relaxation time,
 which can be used to robustly predict barrier-crossing times from simulation
 or experimental trajectory data of general non-Markovian inertial systems
  without the need to extract memory functions.
For a single exponential memory kernel we  give a closed-form expression for the barrier-crossing time, 
 which reproduces the Kramers turnover between the high-friction and high-mass limits 
 as well as the memory turnover from the  intermediate memory acceleration  to the asymptotic  long-memory slow-down regime.
 We also show that non-Markovian systems are singular in the  zero-mass limit, which suggests that 
  the   long-memory barrier-crossing slow-down  reflects the interplay between mass and memory effects. 
 Thus, physically sound models for non-Markovian systems have to include a finite mass.
 
   \end{abstract}

\maketitle

\section{Introduction}

Rare events, that means
chemical reactions, conformational molecular transformation or nucleation in metastable bulk systems,
involve in their time-limiting step  the crossing of a free-energetic  barrier when described in terms of 
low- or one-dimensional reaction coordinates
\cite{Kramers_1940, Visscher1976, Chandler1978, Skinner1978,Hynes1985,Berne_1988}.
The theoretical description of such barrier-crossing phenomena has a long history in physics and chemistry
\cite{Hanggi_1990}. 
Arrhenius showed  that  for a system in contact with a heat bath, 
the typical barrier-crossing time  scales as $\tau_0 \text{e}^{\beta U_0}$, where $\beta=1/(k_\text{B}T)$ is the inverse thermal energy, 
$U_0$ is the height of the  barrier in the free-energy landscape, 
and $\tau_0$ is a prefactor  \cite{Arrhenius_1889}.
That prefactor depends not only on the free-energy profile, as in transition-state theory \cite{Eyring_1935, Laidler_1983}, but also on the dissipative coupling to the environment
\cite{Kramers_1940, Visscher1976, Chandler1978, Skinner1978, Hynes1985,Berne_1988,Allen1980, Trullas1990, Lickert2020}.
Kramers calculated $\tau_0$  in the  Markovian limit,
where environmental dissipation is described by a time-independent friction coefficient $\gamma$, implying infinitely fast environmental relaxation dynamics \cite{Kramers_1940}.
He derived explicit formulas for  $\tau_0$ in the high-friction limit, where inertial effects due to the effective mass 
of the reaction coordinate can be neglected, as well  as in the  low-friction limit, where inertial relaxation effects become dominant.
The crossover between these asymptotic limits, known as the Kramers turnover regime,
was considered in many works \cite{Melnikov_1986,Grabert1988,Best_2006}. 

However, physical  environments do not relax instantaneously. 
The influence of finite environmental relaxation times on reaction coordinate dynamics was explored
 by Zwanzig \cite{Zwanzig_1961} and Mori \cite{Mori_1965}, who used projection techniques to show that the dynamics of a low-dimensional observable defined for 
 a general many-body system is described by the generalized Langevin equation (GLE).
 The GLE  is an  integro-differential equation that generalizes  the  Newtonian equation of motion by explicitly accounting for 
the coupling of the observable to its responsive environment in terms of a non-Markovian  (i.e. time-dependent)
 friction memory kernel $\Gamma(t)$  and a time-dependent force. 
Thus, the GLE correctly accounts for the loss of information when projecting the high-dimensional phase-space dynamics
onto the low-dimensional observable dynamics and constitutes an exact coarse-graining method.
An alternative  strategy to eliminate or reduce  non-Markovian effects is to  use suitable multidimensional reaction coordinates that explicitly account
for solvent degrees of freedom, which basically corresponds to the Markovian-embedding treatment 
of a GLE with multi-exponential memory kernel \cite{Acharya2022}.
Several methods to extract all GLE parameters from time-series data exist 
\cite{straub1987,hijon_morizwanzig_2010,carof_two_2014,jung_iterative_2017,daldrop_external_2017,daldrop_butane_2018, kowalik2019, grogan_data-driven_2020,klippenstein_introducing_2021,vroylandt_likelihood_2022,tepper2024} and using such methods 
the GLE has been applied to  
bond-length vibrations \cite{Tuckerman1993,Gottwald2015,brunig2022a},
dihedral rotations  \cite{Chandler1978,Bagchi_1983,daldrop_butane_2018,dalton2024},
chemical reactions in solvents \cite{Adelman1980,Ciccotti1981,Wolf2020,brunig2022d},
 protein folding \cite{plotkin_non-markovian_1998,lange_collective_2006,satija2019,lee_multi_2019,ayaz2021,dalton2023}
 and also more complex  systems, such as the motion of living organisms as well as genetic, financial and meteorological data
 \cite{schmitt_analyzing_2006, mitterwallner2020, Sollich2020, hassanibesheli2020reconstructing,Metzler_2024,klimek2024,kiefer2025,Dalton_Review_2025}. 

Grote and Hynes (GH) derived a self-consistent equation for barrier-crossing rates in systems with short memory times in the medium- to high-friction regime \cite{Grote_1980}, which was subsequently sucsessfully applied to different barrier-crossing phenomena in liquids \cite{Bagchi1996,Truhlar2000}.
Pollak, Grabert, and H{\"a}nggi subsequently constructed a theory also applicable for long memory times \cite{Pollak_1989,Pollak_1993}, which however does not allow to analytically extract the asymptotic dependence
of barrier-crossing times on the memory time.
Many barrier-crossing rate theories represent the free-energy landscape as an inverted parabola
and thereby neglect relaxation in a free-energy minimum prior to a transition. 
 However, as we show in this paper, this  relaxation  is  important particularly for long memory times, 
  therefore the barrier  free-energy landscape is more realistically modeled as a single-well or 
 double-well potential, relevant, for example, to chemical reactions and protein folding 
 \cite{Berne_1988, Hanggi_1990,Bryngelson_1995, Berezhkovskii_2004, Dudko_2006,  Schuller_2020}. 
In fact,
from numerical simulations it is known that the barrier-crossing time of a non-Markovian one-dimensional reaction coordinate 
with a single exponentially decaying  memory function exhibits a non-monotonic 
dependence on the memory time  \cite{Straub_1986}: keeping the  friction  (i.e. the
integral over the memory function) constant, 
 for intermediate values of the memory time the barrier-crossing time
  is reduced compared to the Markovian limit but for long memory times increases quadratically with the memory time
   \cite{Kappler_2018, Kappler_2019, Lavacchi_2020}.
 The intermediate memory acceleration regime is accurately predicted by Grote-Hynes theory, for the asymptotic  long-memory slow-down behavior
  no systematic  analytically tractable theory is available, a gap that is filled in this paper.

  Starting from the Green function for a general inertial non-Markovian Gaussian reaction coordinate in a harmonic well
  given in Sec. \ref{Green}, 
 we derive in Sec. \ref{Fokker}
 by an exact mapping a generalized Fokker-Planck equation with a time-dependent effective diffusion constant
 that accounts for non-Markovian and inertial  effects. 
 Using the  propagator of the generalized Fokker-Planck operator, absorbing and
reflecting boundary conditions are introduced in  Sec. \ref{Barrier}.
 To first order in a cumulant expansion of the   time-dependent  diffusion constant,  
 we derive in Sec. \ref{Cumulant} an analytical expression for the 
 pre-exponential factor of the Arrhenius prediction  for the barrier-crossing time 
  that involves the integral over the potential-energy correlation function. 
 This expression  can be straightforwardly applied to simulation
 and experimental trajectory data without the need to extract memory functions,
 as is necessary for standard non-Markovian reaction rate theories.
 The influence of different potential shapes on the mean-first passage time is treated in Sec. \ref{Potentials}.
 In Sec. \ref{GLE} we derive a closed-form expression for the barrier-crossing time 
for a single exponential memory kernel,  which reproduces the Kramers turnover between the Markovian high-friction and high-mass limits 
 as well as the memory turnover from the  intermediate memory acceleration, where
 agreement with Grote-Hynes theory is obtained,   to the asymptotic  long-memory slow-down regime.
  In Sec.  \ref{Mass}  we demonstrate
  that non-Markovian systems are singular in the  mass-less limit.
 In   Sec.  \ref{reversibility}  we show that the Langevin equation, when properly derived from
  the GLE in the memory-less limit, preserves time reversibility, as opposed to usual formulations
  of the Langevin equation  in literature. 
  Section \ref{Conclu} summarizes our results in a condensed set of equations and compares our
  predictions of  barrier-crossing times  with previously published simulations results.

\section{Results}

\subsection{  Non-Markovian Green function   } \label{Green}

We consider the  dynamics of a general non-Markovian stochastic process in a harmonic potential well. Neglecting possible non-linear 
friction effects, the process is Gaussian and the two-point positional  joint distribution for the position  
to be $x$ at time $t$ and $x_0$ at time zero  in all generality is given by 
\begin{align}
\label{dist1}
{\cal P}(x,t; x_0) &= \exp\left( - a(t) x^2/2 - b(t) x x_0 - c(t) x_0^2 /2 \right) {\cal N}^{-1},
\end{align}
where $ {\cal N}$ is the normalization constant and $a(t)$, $b(t)$, $c(t)$ are unknown time-dependent functions.
These four unknowns are determined by the following four conditions:

i) For a stationary process (assumed in this work) the joint distribution is symmetric, i.e., 
\begin{align}
\label{dist2}
{\cal P}(x,t; x_0) &= {\cal P}(x_0,t; x),
\end{align}

ii) the joint distribution is  normalized, 
\begin{align}
\label{dist3}
\int_{-\infty}^{\infty} {\rm d}x  \int_{-\infty}^{\infty} {\rm d}x_0  {\cal P}(x,t; x_0) &= 1,
\end{align}

iii) the covariance of the distribution defines the two-point correlation function
\begin{align}
\label{dist4}
\int_{-\infty}^{\infty} {\rm d}x  \int_{-\infty}^{\infty} {\rm d}x_0  x x_0 {\cal P}(x,t; x_0) & \equiv \langle x x_0 \rangle \equiv C(t),
\end{align}

and iv)  the variance of the distribution is denoted as $C_0$
\begin{align}
\label{dist5}
\int_{-\infty}^{\infty} {\rm d}x  \int_{-\infty}^{\infty} {\rm d}x_0  x^2 {\cal P}(x,t; x_0) & = \langle x^2 \rangle =  C(0)\equiv C_0
\end{align}
(note that due to symmetry we have $ \langle x^2 \rangle =  \langle x^2_0 \rangle$). With these four relations, 
the joint distribution can be written in terms of the correlation function $C(t)$  as 
\begin{align}
\label{dist6}
{\cal P}(x,t; x_0) &= \frac{
 \exp\left( 
-\frac{ (x-\bar C(t)x_0)^2}{2 C_0 \left(1-\bar C^2(t)\right) }- \frac{x_0^2}{2 C_0}
 \right) }{
 2 \pi C_0  \left(1-\bar C^2(t)\right) } ,
\end{align}
where we introduced the normalized two-point correlation function as 
\begin{align}
\label{dist7}
\bar C(t)  &= C(t)/C_0,
\end{align}
see Appendix  \ref{AppGreen} for details.
Using the single-point distribution function, which is obtained via marginalization of Eq.(\ref{dist6}) as
\begin{align}
\label{dist8}
{\cal P}(x) &\equiv 
 \int_{-\infty}^{\infty} {\rm d}x_0  {\cal P}(x,t; x_0) =
 \frac{
 \exp\left( 
- \frac{x^2}{2 C_0}  \right) }{
 \left[ 2 \pi/  C_0 \right]^{1/2}} ,
\end{align}
the conditional two-point distribution is obtained as 
\begin{align}
\label{dist9}
{\cal P}(x,t | x_0) &= \frac{{\cal P}(x,t; x_0)}{{\cal P}(x_0)}
= \frac{
 \exp\left( 
-\frac{ (x-\bar C(t)x_0)^2}{2 C_0 \left(1-\bar C^2(t)\right) } \right) }{
 \left[ 2 \pi C_0  \left(1-\bar C^2(t)\right) \right]^{1/2}}.
\end{align}
In fact, the expression Eq.(\ref{dist9})  has been recently derived from the generalized Langevin equation 
using  stochastic path-integral formalism \cite{Netz2025b}, 
which underscores the exactness of the current symmetry-based derivation. 
Since the conditional two-point distribution in Eq.(\ref{dist9}) corresponds to  the distribution  generated by the 
non-Markovian process with the initial probability distribution  ${\cal P}(x,t=0 | x_0) = \delta(x-x_0)$, it
 constitutes  the Green function of the process. 

\subsection{From the non-Markovian Green function to the  generalized Fokker-Planck equation } \label{Fokker}

By taking a partial derivative of the Green function Eq.(\ref{dist9}) with respect to time, we obtain an expression
that can be brought into the  form of a partial differential equation
\begin{align}
\label{FP1}
\frac{ \partial {\cal P}(x,t | x_0) }{\partial t}&=
 - \frac{C_0  \bar C'(t) }{ \bar C(t)} 
\left[  \frac{ \partial}{\partial x} \frac{x}{C_0} +  \frac{ \partial^2}{\partial x^2} \right] {\cal P}(x,t | x_0),
\end{align}
where $\bar C'(t)$ denotes the derivate of $\bar C(t)$,
see Appendix \ref{AppFP} for intermediate steps of the derivation.
The  Green function of the Gaussian process in Eq.(\ref{dist9}) is valid for a harmonic potential of the form 
\begin{align}
\label{FP2}
U(x) &= Kx^2/2,
\end{align}
for which the variance follows for a canonical system at temperature $T$ from the equipartition theorem as $C_0=k_BT/K$,
where $k_B$ is the Boltzmann constant.
With this we can rewrite Eq. (\ref{FP2}) in the form 
\begin{align}
\label{FP3}
\frac{ \partial {\cal P}(x,t | x_0) }{\partial t}&=
D(t)
\left[  \frac{ \partial}{\partial x}\beta U'(x)+  \frac{ \partial^2}{\partial x^2} \right] {\cal P}(x,t | x_0) \nonumber \\
&=  D(t)   \frac{ \partial}{\partial x}   e^{ -  \beta U(x) }
 \frac{ \partial}{\partial x}   e^{  \beta U(x) }  {\cal P}(x,t | x_0),
\end{align}
where we used $\beta=1/(k_BT)$ and introduced the time-dependent diffusivity 
\begin{align}
\label{FP4}
D(t)&=
 - \frac{C_0  \bar C'(t) }{ \bar C(t)}=
 - \frac{C_0  C'(t) }{C(t)}.
\end{align}
By writing the probability distribution as a convolution of the Green function ${\cal P}(x,t | x_0)$ with the 
initial distribution ${\cal P}_0 (x_0)$ as
\begin{align}
\label{FP5}
{\cal P}(x,t ) &= \int_{-\infty}^{\infty} {\rm d} x_0 
{\cal P}_0 (x_0) {\cal P}(x,t | x_0),
\end{align}
we obtain from Eq. (\ref{FP3}) an equation for the distribution function 
\begin{align}
\label{FP6}
\frac{ \partial {\cal P}(x,t ) }{\partial t}&=
 D(t)   \frac{ \partial}{\partial x}   e^{ -  \beta U(x) }
 \frac{ \partial}{\partial x}   e^{  \beta U(x) }  {\cal P}(x,t )
\end{align}
with the initial condition $ {\cal P}(x,0 )=   {\cal P}_0 (x)$.
This equation is identical to the standard Fokker-Planck equation for an overdamped Markovian reaction coordinate 
 \cite{zwanzig_nonequilibrium_2001}
except that the diffusivity $D(t)$ is time-dependent. This time-dependency takes care of inertial and non-Markovian effects and
therefore contains the crucial physics we  are interested in.

\subsection{Barrier-crossing times from the generalized Fokker-Planck equation} \label{Barrier}

In the following we extend the method to calculate mean-first-passage times
from the standard Fokker-Planck equation with a time-independent diffusivity \cite{zwanzig_nonequilibrium_2001} to the case
of a  time-dependent diffusivity.
The solution of  the generalized Fokker-Planck  equation for the Green function Eq. (\ref{FP3}) 
can  be written formally as an operator exponential as

\begin{align}
\label{MFP1}
{\cal P}(x,t | x_0)  &= e^{T(t) {\cal D} (x)} {\cal P}(x,0 | x_0),
\end{align}
where we defined  the diffusive Fokker-Planck propagator as 
\begin{align}
\label{MFP2}
 {\cal D} (x) &\equiv
 \frac{ \partial}{\partial x}   e^{ -  \beta U(x) }
 \frac{ \partial}{\partial x}   e^{  \beta U(x) }
\end{align}
and the running integral over the diffusivity as 
\begin{align}
\label{MFP3}
T(t) &\equiv \int_0^t {\rm d}s D(s).
\end{align}
Note that Eq. (\ref{MFP1}), which employs 
  operator-exponential formalism \cite{zwanzig_nonequilibrium_2001},
 is the same Green function as given in  
Eq. (\ref{dist9}), but is more  useful for imposing spatial boundary conditions as needed for 
calculating the  mean-first passage time. 
Using that ${\cal P}(x,0 | x_0) =\delta(x-x_0)$, we can write the Green function in 
Eq. (\ref{MFP1}) in
symmetric form as
\begin{align}
\label{MFP4}
{\cal P}(x,t | x_0)  &= \int_{-\infty}^{\infty} {\rm d} \tilde x 
 \delta(\tilde x-x) e^{T(t) {\cal D} (\tilde x)} \delta(\tilde x-x_0) \nonumber \\ 
 &= \int_{-\infty}^{\infty} {\rm d} \tilde x 
 \delta(\tilde x-x_0) e^{T(t) {\cal D}^\dagger (\tilde x)} \delta(\tilde x-x) \nonumber \\ 
  &= e^{T(t) {\cal D}^\dagger (x_0)} {\cal P}(x,0 | x_0),
 \end{align}
where we defined the adjunct Fokker-Planck operator as 
\begin{align}
\label{MFP5}
 {\cal D} ^\dagger (x) &\equiv
 e^{  \beta U(x)}  \frac{ \partial}{\partial x}   e^{ -  \beta U(x) }
 \frac{ \partial}{\partial x}    .
\end{align}
From Eqs. (\ref{MFP1})  and (\ref{MFP4}) it transpires that the Green function  satisfies the 
normal Fokker-Planck Eq. (\ref{FP3}) as well as the
adjunct (backward) Fokker-Planck equation, i.e. 
\begin{align}
\label{MFP6}
\frac{ \partial {\cal P}(x,t | x_0) }{\partial t}&=
D(t) {\cal D} (x) {\cal P}(x,t | x_0) \nonumber \\ 
&=   D(t) {\cal D}^\dagger (x_0) {\cal P}(x,t | x_0).
\end{align}
We define the survival probability in the domain between a reflecting boundary condition at $x_{\rm ref}$
and an absorbing boundary condition at the position  $x_f> x_{\rm ref} $ as 
\begin{align}
\label{MFP7}
S(x_0,t) &= \int_{x_{\rm ref}}^{x_f} {\rm d}  x {\cal P}(x,t | x_0),
\end{align}
which has the limits $S(x_0,0)=1$ and  $S(x_0,\infty)=0$ for $x_{\rm ref}< x_0<x_f$.
The first-passage (or escape) distribution follows  from the time derivative of the survival probability as 
\begin{align}
\label{MFP8}
k(x_0,t) &\equiv -\frac{\partial}{\partial t} S(x_0,t) = 
- \int_{x_{\rm ref}}^{x_f} {\rm d}  x \frac{\partial}{\partial t}  {\cal P}(x,t | x_0) \nonumber \\ 
& = -  \int_{x_{\rm ref}}^{x_f} {\rm d}  x D(t) {\cal D} (x) {\cal P}(x,t | x_0) \nonumber \\ 
&= - D(t) e^{ -  \beta U(x_f) }
 \frac{ \partial}{\partial x_f}   e^{  \beta U(x_f) } {\cal P}(x_f,t | x_0),
\end{align}
where we employed Eq. (\ref{MFP6}) and used
 in the last step that the probability  flux at $x_{\rm ref}$ vanishes.
Applying the operator expression for  $k(x_0,t)$ in Eq. (\ref{MFP8}) 
onto the adjunct Fokker-Planck Eq. (\ref{MFP6}), we obtain 
\begin{align}
\label{MFP9}
\frac{ \partial k(x_0,t) }{\partial t}
=   D(t) {\cal D}^\dagger (x_0) k(x_0,t).
\end{align}
From this partial differential equation for the first-passage distribution
 we can recursively determine moments of the product of the
 first-passage distribution  and the diffusivity defined as
\begin{align}
\label{MFP10}
k_n(x_0) \equiv  \int_0^\infty  {\rm d}t  t^n D(t) k(x_0,t)/D_0,
\end{align}
where $D_0$ is  the diffusivity in a suitably defined  limit, as will be discussed  further below.
For the first moment we obtain  from Eq. (\ref{MFP9})
\begin{align}
\label{MFP11}
{\cal D}^\dagger (x_0) k_1(x_0) &= \int_0^\infty {\rm d}t t \frac{\partial}{\partial t} k(x_0,t)/D_0
\nonumber \\
&= -   \int_0^\infty {\rm d}t  k(x_0,t)/D_0 = -1/D_0,
\end{align}
where we used that $k(x_0,\infty)=0$ and that the time  integral over $k(x_0,t)$ is normalized to unity.
This equation  can be inverted and yields
\begin{align}
\label{MFP12}
k_1(x_0)  = D_0^{-1}   \int_{x_0}^{x_f} {\rm d}x e^{ \beta U(x) }
 \int_{x_{\rm ref}}^{x} {\rm d}x' e^{ -  \beta U(x') }.
 \end{align}
Note that from the definition Eq. (\ref{MFP10}) it transpires that 
$k_1(x_0)$ only corresponds to the mean-first passage time  $\tau_{\rm mfp}$,
 which is the average time to
reach the final position $x_f$ for the first time when  starting from  $x_0$,
if the diffusivity $D(t)$ is  time-independent and given by $D_0$.
If, on the other hand,   the diffusivity $D(t)$  is time dependent,
$k_1(x_0)$ does not equal   $\tau_{\rm mfp}$, in which case 
 there are different ways of how to proceed. Using that the 
first-passage distribution  for high barriers  decays single-exponentially as 
$k(x_0,t)\simeq e^{-t/\tau_{\rm mfp}}/ \tau_{\rm mfp}$, 
$ \tau_{\rm mfp}$ is  determined by the equation following from combining   
Eq. (\ref{MFP10})  for $n=1$  and 
 given $D(t)$ with   Eq. (\ref{MFP12}). 
A numerical solution of the resulting equation 
 is possible if the two-point correlation function $C(t) $ is known reliably  and  the diffusivity profile $D(t)$ 
can via Eq. (\ref{FP4}) be accurately calculated. In typical simulation or experimental scenarios, however,  the
correlation function becomes noisy in the long-time limit and thus $D(t \rightarrow \infty)$ cannot be reliably estimated.
An additional problem ist that $D(t)$  for inertial and non-Markovian systems typically oscillates in the long-time limit and thus
the limit   $D(t \rightarrow \infty)$ is not uniquely determined, as will be seen later.
In the following section we will introduce a robust expansion method for $D(t)$, which allows us to replace
$D(t)$ by a constant $D_0$ to leading order in a systematic cumulant expansion.

\subsection{Cumulant expansion of the time-dependent diffusivity $D(t)$} \label{Cumulant}

The key observation is that for a single-exponential correlation function $C(t)$, the
diffusivity $D(t)$ defined in Eq. (\ref{FP4}) becomes time-independent, i.e. $D(t)=D_0$, and thus the
mean-first passage time  equals the  first moment   in Eq. (\ref{MFP12}), i.e.   $\tau_{\rm mfp} = k_1(x_0)$.
This is a non-trivial observation, as the correlation function of non-Markovian and inertial systems do for specific parameter combinations
indeed decay single-exponentially, as will be explicitly demonstrated below.
The idea pursued in this section is to characterize deviations of $C(t)$  from a single-exponential  function
by a systematic cumulant expansion. 

To proceed, we rewrite Eq. (\ref{FP4}) as 
\begin{align}
\label{FP7}
D(t)&= - \frac{C_0  \bar C'(t) }{ \bar C(t)}=
 - \frac{C_0   }{p  }  \frac{ {\rm d}{\bar C}^p(t)/{\rm d}t }{ {\bar C}^p(t)},
\end{align}
where $p$ is an arbitrary positive integer number that will be discussed and chosen later.
We furthermore define the single-sided function 
\begin{align}
\label{FP8}
\phi(t) &\equiv   - \theta(t)  \frac{{\rm d}{\bar C}^p(t) }{ {\rm d}t},
\end{align}
where $ \theta(t)$ defines the Heaviside function.
The Fourier transform of $\phi(t)$ can be written as a power series in frequency according to 
\begin{align}
\label{FP9}
\tilde \phi(\omega) &= \int_{-\infty}^\infty {\rm d}t e^{-\imath \omega t}  \phi(t) \nonumber \\
&= 1-\imath \omega \sum_{n=0}^\infty \frac{(-\imath \omega)^n}{n!} M_n,
\end{align}
where we defined moments of the $p$-th power of the correlation function as
\begin{align}
\label{FP10}
M_n = \int_0^\infty {\rm d}t t^n {\bar C}^p(t).
\end{align}
As the only requirement, we assume all higher moments $M_n$ to exist, which is satisfied for exponentially
decaying correlation functions, which is the normal scenario. 
Note that by virtue  of the presence of the derivative in the definition of $\phi(t)$  in Eq. (\ref{FP8}) and the
single-sidedness of  $\phi(t)$,  the power series in   Eq. (\ref{FP9}) starts with unity, which is crucial.
In order to make the  back-Fourier transform integrable, we define the inverse power series
\begin{align}
\label{FP11}
\tilde \phi(\omega) &= 
\left[ 1+\imath \omega \sum_{n=0}^\infty  \omega^n  B_n \right]^{-1}
\end{align}
with coefficients $B_n$ that follow uniquely from the set of moments $M_m$ with $m \leq  n$ by standard
power-series inversion. To illustrate this, we write out the first two terms explicitly
\begin{align}
\label{FP12}
\tilde \phi(\omega) &= 1- \imath \omega M_0 - \omega^2 M_1 + \cdots \nonumber \\ 
&= \left[ 1+\imath \omega M_0 + \omega^2 \left( M_1 - M_0^2 \right)  + \cdots  \right]^{-1}.
\end{align}
It transpires that $\tilde \phi(\omega)$, when written as an  
inverse power series in $\omega$, 
contains multiple poles that  will lead upon back-Fourier-transforming to 
a sum of exponentially decaying functions. To linear order in $ \omega $ we find 
$\tilde \phi(\omega) \simeq  1/(1+\imath \omega M_0)$, which leads to the single exponential
\begin{align}
\label{FP13}
 \phi(t) &\simeq  \theta(t) e^{-t/M_0}/M_0.
 \end{align}
Since for a single exponential correlation function $\bar C(t)= e^{-t/\tau}$ one has $M_1 = M_0^2$,
one finds that in this case  the second-order correction in $\omega$ in Eq. (\ref{FP12}) vanishes (and so do all higher-order corrections). 
In essence, the cumulant expansion allows to write down a functional expansion in terms of  deviations of the
correlation function $\bar C(t)$ from a single-exponential form.
Note that the first-order result  in Eq. (\ref{FP13}) does not   reproduce the long-time asymptotic decay of 
$\phi(t)$ in case there are competing exponentially decaying contributions, this is particularly true 
 for correlation functions that asymptotically oscillate, as is the case for highly  inertial or highly non-Markovian systems,
 which will be illustrated below.

We so far have considered the numerator of Eq. (\ref{FP7}). The denominator of Eq. (\ref{FP7}) can for $t>0$
be expressed in terms of 
the function $\phi(t)$ as 
\begin{align}
\label{FP14}
 {\bar C}^p(t) &= -  \int_t^\infty {\rm d}s \frac{{\rm d}  {\bar C}^p(s)}{ {\rm d} s} =  \int_t^\infty {\rm d}s \phi(s)
 \simeq e^{-t/M_0},
 \end{align}
where in the last step we used the first-order  result from Eq. (\ref{FP13}).
By combining the results in Eqs. (\ref{FP13}) and (\ref{FP14}) we obtain from Eq. (\ref{FP7}) for the diffusivity 
 to  leading order in the cumulant expansion 
\begin{align}
\label{FP15}
D(t)&\simeq D_0 \equiv   \frac{C_0  }{ \tau_{\rm rel}} =  \frac{C_0  }{ p M_0 } ,
\end{align}
where we defined  the relaxation time as  $\tau_{\rm rel}\equiv C_0/D_0$,
for which we obtain to leading order in the cumulant expansion $\tau_{\rm rel}= p M_0$.
For $p=2$, which is a natural choice as we will show below,  the relaxation time follows
from Eq. (\ref{FP10}) as
the integral over the squared correlation function
\begin{align}
\label{FP16}
\tau_{\rm rel} & =2  \int_0^\infty {\rm d}t {\bar C}^2(t)
=\frac{2}{C_0^2}  \int_0^\infty {\rm d}t { C}^2(t) \nonumber \\
& =\frac{ K^2}{(k_BT)^2}  \int_0^\infty {\rm d}t \left( 
\langle x^2(t) x^2(0) \rangle - \langle x^2 \rangle^2 \right),
\end{align}
where in the last line we used Wick's theorem to show the correspondence 
to the  potential-energy correlation function (in the following  simply called the energy correlation function). 
Note that higher-order corrections can be calculated by carrying the cumulant expansion in 
Eq. (\ref{FP12}) to higher order  in $\omega$, in which case   the diffusivity $D(t)$ becomes
time-dependent and the  mean-first-passage time $\tau_{\rm mfp}$
does not  equal the  first moment   in Eq. (\ref{MFP12}), which is not pursued in this paper.

\begin{figure*}	
	\centering
	\includegraphics[width=18cm]{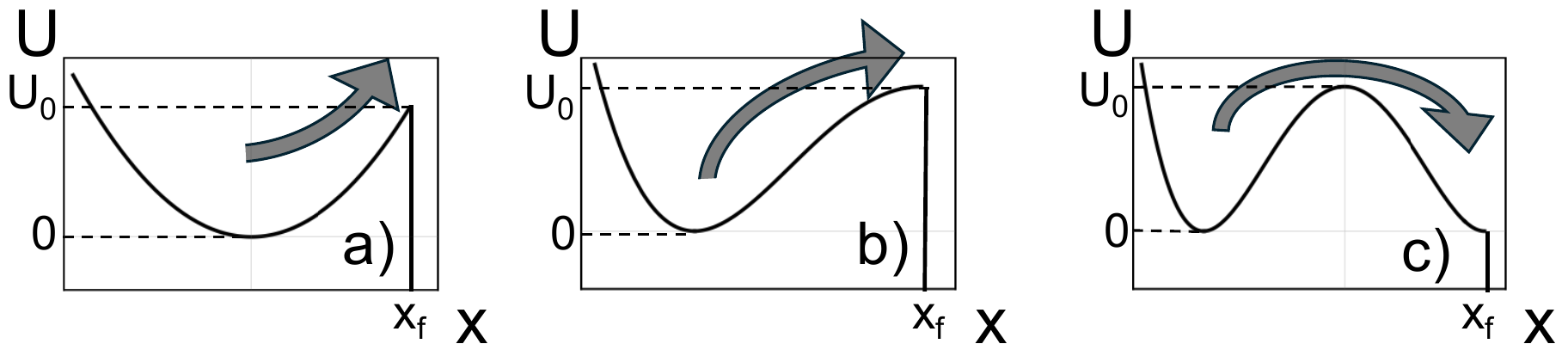}
\caption{ {   Illustration of different potential shapes considered for the calculation of the mean-first passage time
according to Eq. (\ref{MFP12}). The adsorbing boundary condition is indicated by an abrupt decrease of the potential.
a) Harmonic potential. b) Absorbing potential is located on the top of a barrier with vanishing slope. c) Absorbing boundary condition
is located to  the right of the barrier.   }
		}
	\label{fig1}
\end{figure*}

To show that $p=2$ is a rather natural choice, we consider the  average over the time-dependent 
diffusivity defined in Eq. (\ref{FP7}) with a weight function $w(t)$, which we take to be given 
by the $p$-th power of the rescaled correlation function, $w(t)= {\bar C}^p(t)$, according to 
\begin{align}
\label{FP17}
\bar D &=  \frac{ \int_0^\infty {\rm d}t D(t) w(t)}{  \int_0^\infty {\rm d}t  w(t)} = 
- C_0  \frac{ \int_0^\infty {\rm d}t  {\bar C}'(t)  {\bar C}^{p-1}(t) }{  \int_0^\infty {\rm d}t  {\bar C}^p(t) } =  \frac{C_0  }{ p M_0 }.
 \end{align}
We see that we reproduce exactly the result  for $D_0$ 
in Eq. (\ref{FP15}), which means that the leading order of the cumulant
expansion corresponds to the  average over $D(t$) with a weight function proportional 
to $ {\bar C}^p(t)$. This makes  sense, since an effective time-independent diffusivity should reflect
the time-dependent diffusivity in  the  time range   that is relevant for the escape process. 
Since the idea of a weighting function implies a positive definite function and the correlation function ${\bar C}(t)$
can be negative even in the asymptotic long-time limit, 
an even exponent $p$ makes sense. Furthermore, for $p=2$ the relaxation time
defined in Eq. (\ref{FP13}) corresponds to the energy relaxation time, which makes also  intuitive
sense, as barrier-crossing involves energy relaxation.  Thus, we will use $p=2$ in all further calculations
and confirm the validity of this choice by comparison with  previously published mean-first passage times
 determined from simulations of barrier crossing dynamics in double-well potentials.

\subsection{Explicit predictions for different potential shapes} \label{Potentials}

In the high-barrier limit
the double integral in Eq. (\ref{MFP12}) can be solved by saddle-point methods \cite{zwanzig_nonequilibrium_2001}.
In Fig. \ref{fig1} the absorbing boundary condition
is for a few different potentials  illustrated by an abrupt decrease of the potential, indicating that a particle  coming from the left
that crosses the absorbing boundary condition  will not be able to return back to the left.
Since the Green function is calculated using Gaussian methods, our results are exact (except for  the cumulant expansion 
used to approximate the diffusivity $D(t)$ as a constant $D_0$) 
 for the harmonic potential shown in Fig. \ref{fig1}a). 
Our methods are   expected to be also rather accurate for the mean-first passage time
 to reach the barrier top, 
illustrated in Fig. \ref{fig1}b, since the difference to a harmonic potential in a) is rather small.
The difference of the mean-first passage time between crossing over the barrier to the other side of
a double-well potential, as illustrated  in Fig. \ref{fig1}c),
and reaching the barrier-top, as illustrated  in Fig. \ref{fig1}b),
 however, is not expected to be represented accurately  by our method for strongly non-Markovian systems.

To first order in the cumulant expansion introduced in Section \ref{Cumulant} and using the saddle-point approximation,
the mean-first passage time to reach the sharp-kink barrier of a harmonic potential (scenario in Fig. \ref{fig1}a)  follows
from Eq. (\ref{MFP12}) as 
\begin{align}
\label{pot1}
\tau_{\rm mfp}^a   = \sqrt{  \frac{ \pi}{\beta U_0 } }   \frac{ e^{ \beta U_0 }}
{D_0   \beta K } 
= \tau_{\rm rel} \sqrt{  \frac{ \pi}{\beta U_0 } }    e^{ \beta U_0 },
 \end{align}
the mean-first passage time to reach the flat barrier top (scenario in Fig. \ref{fig1}b)  reads 
\begin{align}
\label{pot2}
\tau_{\rm mfp}^b = \frac{ \pi e^{ \beta U_0 }}
{D_0   \beta \sqrt{K K_{\rm bar}} }=
\pi \tau_{\rm rel}   \left( \frac{K }{ K_{\rm bar}}\right)^{1/2}     e^{ \beta U_0 },
 \end{align}
and the mean-first passage time to reach over the  barrier to the other side (scenario in Fig. \ref{fig1}c) is given by
\begin{align}
\label{pot3}
\tau_{\rm mfp}^c = \frac{2 \pi e^{ \beta U_0 }}{D_0   \beta \sqrt{K K_{\rm bar}} }
=2 \pi \tau_{\rm rel}   \left( \frac{K }{ K_{\rm bar}}\right)^{1/2}   e^{ \beta U_0 },
 \end{align}
where we used the definition of $\tau_{\rm rel}$ in Eq. (\ref{FP15}), the 
equipartition result for a harmonic potential $C_0= k_BT / K$
and where $K_{\rm bar}$ is  the absolute value of the potential curvature at the barrier top.
For details of the saddle-point approximation see Appendix \ref{AppMFPT}.
 The  energy relaxation time
  $\tau_{\rm rel}$ can be determined for model systems or  from experimental or simulation trajectories 
   by the integral over the squared correlation function (determined in the free-energy well) 
    according to Eq. (\ref{FP16}).
We see that the different potential shapes shown in Fig. \ref{fig1} lead to distinct pre-exponential factors
and a dependence on the barrier height $U_0$ for the case of a harmonic potential with an absorbing cusp in Fig. \ref{fig1}a.
For potentials with rather similar curvatures $K$ and $K_{b\rm ar}$ in the well and at the barrier the 
factor $ \sqrt{K / K_{\rm bar}}$ is expected to be rather close to unity.

\subsection{ Explicit results for relaxation times} \label{GLE}

We consider the Mori GLE  for the reaction coordinate (also called the  position for simplicity)   $x(t)$
\begin{align}
\label{GLE1}
m \ddot x(t) &= -K x(t) - \int_{-\infty}^{\infty} {\rm d} s \Gamma(s) \dot x(t-s) + F(t),
 \end{align}
where $m$ is the effective mass, defined 
as $m= k_BT/\langle \dot x^2 \rangle$, $K$ is the stiffness of the harmonic potential,
  the memory kernel $\Gamma(t)$  is here (for simplifying the Fourier analysis) 
  defined to be a single-sided function, i.e. $\Gamma(t)=0$ for $t<0$,
and we have moved the projection time to minus infinity, such that the integration  of the non-Markovian
friction extends over the entire time domain (see Section \ref{reversibility} for a discussion of how 
Eq. (\ref{GLE1})  follows from projection theory).
The  autocorrelation function of the orthogonal force  is for a time-independent Hamiltonian given by 
\begin{align}
\label{GLE2}
\langle F(t) F(t') \rangle  &= k_BT  \Gamma(|t-t'|),
 \end{align}
where  averages in this Section  are  taken over the orthogonal-force distribution. 
Note that the Mori GLE in Eq. (\ref{GLE1}) is an exact equation of motion for a general non-linear reaction coordinate, unless 
one approximates the orthogonal force $F(t)$ as Gaussian, as we do in this paper and as is appropriate for a Gaussian system.
Fourier-transformation of Eq. (\ref{GLE1}) leads to 
\begin{align}
\label{GLE3}
- m \omega^2  \tilde x(\omega) &= -K \tilde x(\omega) - \imath \omega \tilde \Gamma(\omega)   \tilde x(\omega) + \tilde F(\omega),
 \end{align}
from which the response of the position to the orthogonal force follows as
\begin{align}
\label{GLE4}
 \tilde x(\omega)  = \tilde \chi(\omega) \tilde F(\omega) 
 \end{align}
with the response function  given by
\begin{align}
\label{GLE5}
 \tilde \chi(\omega) = \left[ K  - m \omega^2 +  \imath \omega \tilde \Gamma(\omega) \right]^{-1}.
 \end{align}
From  Eqs. (\ref{GLE2}) and  (\ref{GLE4}) the Fourier transform of the correlation function $C(t)= \langle x(s) x(s+t) \rangle$
 follows as
\begin{align}
\label{GLE6}
 \frac{\tilde C(\omega)}{k_BT}  &= \left[  \tilde \Gamma(\omega) + \tilde \Gamma(-\omega)   \right]  
 \tilde \chi(\omega)  \tilde \chi(-\omega) 
 \nonumber \\
 &=  \frac{  \tilde \Gamma(\omega) + \tilde \Gamma(-\omega)  }{ \left| K  - m \omega^2 +  \imath \omega \tilde \Gamma(\omega) \right|^2} \nonumber \\
 &= \frac{\tilde \chi(-\omega)}{\imath \omega} -   \frac{\tilde \chi(\omega)}{\imath \omega},
 \end{align}
where the last equation follows directly from the definition of the response function in Eqs. (\ref{GLE5})
and where we used that $\chi(t)$ and $\Gamma(t)$ are real functions in the time domain.
We define the derivative of the correlation function as 
\begin{align}
\label{GLE7}
  C_{xv}(t) &  \equiv \frac{{\rm d}C(t)}{ {\rm d}t}  =  \frac{{\rm d} }{ {\rm d}t}
   \int \frac{{\rm d}\omega}{2\pi} e^{\imath \omega t}  \tilde C(\omega) \nonumber \\
&  = \int \frac{{\rm d}\omega}{2\pi} e^{\imath \omega t} \imath \omega  \tilde C(\omega),
 \end{align}
from which transpires  that $ \tilde C_{xv}(\omega)= \imath \omega \tilde C(\omega)$ and thus from Eqs. (\ref{GLE6})
\begin{align}
\label{GLE8}
 \frac{\tilde C_{xv}(\omega)}{k_BT}  = 
 \tilde \chi(-\omega) -  \tilde \chi(\omega).
 \end{align}
The time-domain  response function $\chi(t)$  is a single-sided function,
therefore the Fourier transform of the single-sided correlation function 
$C_{xv}^+(t)\equiv \theta(t) C_{xv}(t)$ is, if the response function $\chi(t)$ 
 is regular  in the limit $t\rightarrow0$ (we will come back to this important 
 condition when we discuss the massless non-Markovian scenario in Sec. \ref{Mass})  given by
\begin{align}
\label{GLE9}
 \frac{\tilde C^+_{xv}(\omega)}{k_BT}  =  -  \tilde \chi(\omega).
 \end{align}
This is a  useful result for analytical calculations,
as the response function $ \tilde \chi(\omega)$ has the  minimal
pole structure. From the time-domain function
$C_{xv}^+(t)$ the single-sided correlation function $C^+(t) \equiv \theta(t) C(t)$ 
 is obtained by straightforward integration  as 
\begin{align}
\label{GLE10}
  C^+(t) &  =  -\theta(t) \int_t^\infty {\rm d}s  C^+_{xv}(s)
  = k_BT \theta(t)  \int_t^\infty {\rm d}s  \chi(s),
 \end{align}
which can then be used to calculate the relaxation time $\tau_{\rm rel}$   according to Eq. (\ref{FP16}).
Note that the time-domain result in Eq. (\ref{GLE10}) is equivalent to the Fourier-domain result
\begin{align}
\label{GLE10b}
  \tilde C^+(\omega) & 
  = \frac{k_BT}{\imath \omega} \left( \tilde  \chi(0) -  \tilde  \chi(\omega) \right).
 \end{align}
In the following we will consider a single-exponential memory kernel
\begin{align}
\label{GLE11}
  \Gamma(t) &  =  \theta(t) \gamma e^{-t\tau}/\tau
   \end{align}
with a memory time $\tau$, 
whose Fourier transform is given by 
\begin{align}
\label{GLE12}
\tilde  \Gamma(\omega) &  = \frac{  \gamma }{1+\imath \tau \omega}.
   \end{align}
%

\subsubsection{Two-pole analysis for Markovian system: Kramers turnover} \label{2pole}

\begin{figure}	
	\centering
	\includegraphics[width=8cm]{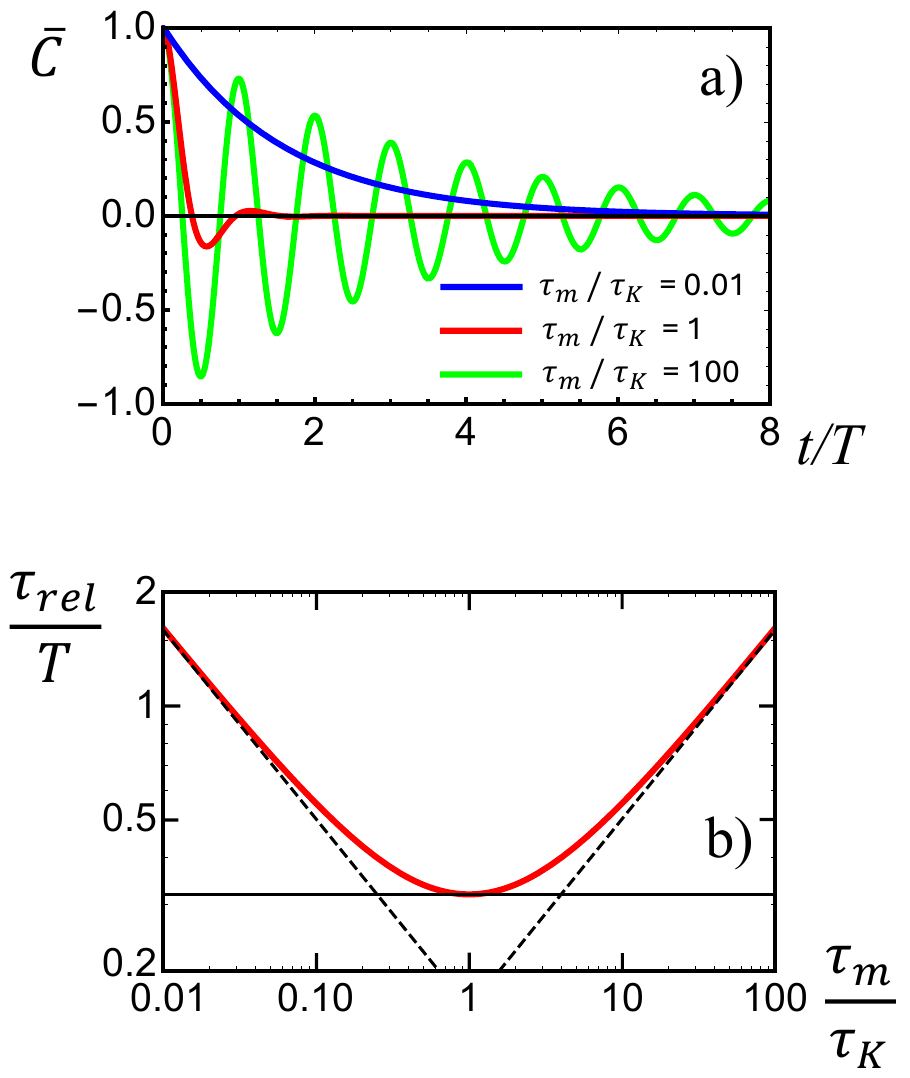}
\caption{ Analytical results in the Markovian two-pole limit corresponding to vanishing memory time $\tau=0$.
a)  Rescaled correlation function $\bar C(t)=C(t)/C_0$ according to Eqs. (\ref{GLE15}) and  (\ref{GLE16})
and using the results for the poles in Eq. (\ref{GLE18}) for three different rescaled masses $\tau_m/\tau_K=mK/\gamma^2$, where
$\tau_m=m/\gamma$ is the inertial relaxation time and $\tau_K=\gamma/K$ is the overdamped  relaxation time. 
Time $t$  is  rescaled by the harmonic oscillation period in the frictionless limit  $T=2 \pi \sqrt{m/K}$. 
b)  Rescaled relaxation  time $\tau_{\rm rel} /T$ according to Eqs. (\ref{GLE20})  
as function of the rescaled mass $\tau_m/\tau_K=mK/\gamma^2$. The asymptotic scalings in the high-mass limit (high $\tau_m/\tau_K$)
$\tau_{\rm rel} \sim m/\gamma$ and in the high-friction limit (small $\tau_m/\tau_K$)
$\tau_{\rm rel} \sim \gamma/K$  are indicated by straight broken lines.
At the minimum at $\tau_m/\tau_K=mK/\gamma^2=1$ the relaxation time is given by $\tau_{\rm rel} = T/\pi$, 
which corresponds to 
the   transition-state-theory prediction. 
		}
	\label{fig2}
\end{figure}

We first consider the Markov limit, obtained for vanishing memory time $\tau \rightarrow 0$, in which case
the Fourier-transformed memory kernel in Eq. (\ref{GLE12}) reads $\tilde  \Gamma(\omega) = \gamma$
and we are thus left with the damped harmonic oscillator model.
The response function in Eq. (\ref{GLE5}) simplifies as 
\begin{align}
\label{GLE13}
 \tilde \chi(\omega) = \frac{1}{ K  - m \omega^2 +  \imath  \gamma \omega}
 = - \frac{1}{ m( \omega - \omega_1)(\omega - \omega_2)}
 \end{align}
with the two poles  given by
\begin{align}
\label{GLE18}
 \omega_{1,2} = \frac{ \imath  \gamma}{2m} \pm 
 \left( \frac{K}{m} - \frac{\gamma^2}{4 m^2} \right)^{1/2}.
 \end{align}
We note that the two poles coincide if the argument of the square root vanishes, which happens at the
crossover between the overdamped and  underdamped regimes and  in which case the correlation function
decays single-exponentially, as announced earlier. 
The response function in the time domain  follows from residual calculus as 
\begin{align}
\label{GLE14}
  \chi(t) = - \frac{ \theta(t)}{m}  \left( \frac{\imath e^{\imath \omega_1 t}}{  \omega_1 - \omega_2}
  + \frac{\imath e^{\imath \omega_2 t}}{ \omega_2 - \omega_1} \right)
   \end{align}
with $\chi(t=0)=0$, i.e., there is no instantaneous response for finite mass. 
The  correlation function follows for positive times from Eq. (\ref{GLE10}) as
\begin{align}
\label{GLE15}
 C^+(t) =  \frac{ \theta(t) k_BT }{m}  \left(  \frac{e^{\imath \omega_1 t}}{  \omega_1( \omega_1 - \omega_2)}
  + \frac{  e^{\imath \omega_2 t}}{  \omega_2 ( \omega_2 - \omega_1)} \right),
   \end{align}
from which the variance follows as 
\begin{align}
\label{GLE16}
 C_0=2 C^+(0) =  - \frac{ k_BT }{m   \omega_1 \omega_2}
= \frac{k_BT}{K},
   \end{align}
in agreement with the equipartition theorem and
where we used the convention for the Heaviside function that $\theta(0)=1/2$.
In Fig. \ref{fig2}a) we plot  $\bar C(t)=C(t)/C_0$
 for three different rescaled masses $\tau_m/\tau_K=mK/\gamma^2$
as a function of time $t$  rescaled by the harmonic oscillation period $T=2 \pi \sqrt{m/K}$,
where $\tau_m=m/\gamma$ is the inertial relaxation time and $\tau_K=\gamma/K$ is the overdamped relaxation time. 
We see that for small mass, $\tau_m/\tau_K=0.01$ (blue line), the correlation function decays monotonically,  for 
large mass, $\tau_m/\tau_K=100$ (green line), long-range oscillations are visible, while for 
intermediate mass, $\tau_m/\tau_K=1$ (red line), the decay is fastest and 
accompanied by only a few oscillations. It transpires that 
the relaxation  time Eq. (\ref{FP16}) defined by the decay of the squared correlation function
 is minimal for intermediate mass (or intermediate friction). 

From Eqs. (\ref{GLE15}) and  (\ref{GLE16}) the energy  relaxation time defined in  Eq. (\ref{FP16}) follows
as 
\begin{align}
\label{GLE17}
\tau_{\rm rel} & = - \frac{ ( \omega_1 + \omega_2)^2 +  \omega_1 \omega_2}{\imath  \omega_1 \omega_2 ( \omega_1 + \omega_2)}.
\end{align}
By inserting  Eq. (\ref{GLE18}) into Eq. (\ref{GLE17}) we obtain the relaxation time as
\begin{align}
\label{GLE20}
\tau_{\rm rel} & = \frac{\gamma}{K} + \frac{m}{\gamma},
\end{align}
which is shown in Fig. \ref{fig2}b) and transparently displays  the Kramers turnover: for  large friction coefficient $\gamma$
(small rescaled mass  $\tau_m/\tau_K=mK/\gamma^2$)
the relaxation time  is linear in  $\gamma$, and thus also the barrier-crossing time increases linear in  $\gamma$,
 as follows from Eq. (\ref{MFP12}),
 while for small $\gamma$ (large  rescaled mass  $\tau_m/\tau_K=mK/\gamma^2$) 
the relaxation time and thus  the barrier-crossing time  is  proportional to  $1/\gamma$.
In between these two limits the relaxation time exhibits a minimum for a friction given by $\gamma^2 = mK$,
at this minimum the relaxation time is given by $\tau_{\rm rel}  = 2 \sqrt{m/K}$, which apart from a numerical factor
is equivalent to the transitions-state-theory prediction, as is explained in Appendix  \ref{AppGH}.

The asymptotic behaviors  captured by  Eq. (\ref{GLE20}) follows from pole analysis:
In the high-friction limit $\gamma^2 \gg mK$ we obtain  from Eq. (\ref{GLE18})
$\omega_1 \simeq \imath K/\gamma$ and $\omega_2 \simeq \imath \gamma/ m$.
Since  $\omega_1 \ll \omega_2$ we conclude that $\omega_1$ describes the asymptotic long-time decay and
thus is the dominant pole,  we see from Eq. (\ref{GLE20}) that the relaxation time in the high-friction limit indeed  is given by
$\tau_{\rm rel} \simeq \imath / \omega_1$. 
Since in the overdamped limit $m \rightarrow 0 $  the only pole is $\omega_1$ (as follows directly from Eq. (\ref{GLE13})),
it transpires that the correlation function $C(t)$  in this limit  decays single-exponentially, 
thus the diffusivity in Eq. (\ref{FP7}) is time-independent and our approach is exact since no cumulant expansion is needed; noteworthily, our approach is also exact
at the crossover friction $\gamma^2 =4 mK$, since also here $C(t)$ decays single-exponentially. 

In the low-friction limit $\gamma^2 \ll mK$ we obtain  from Eq. (\ref{GLE18})
$\omega_{1,2} \simeq \imath \gamma/ (2 m) \pm \sqrt{K/m}$. The inverse  imaginary part of $\omega_{1,2}$ sets the decay time,
which diverges with increasing mass and dominates the relaxation time in Eq. (\ref{GLE20}), as visualized in Fig. \ref{fig2}. 
Note that in the  underdamped limit $m \rightarrow \infty $
the decay of the correlation function does not become single-exponential and thus our cumulant-expansion approach is approximate.

\subsubsection{Three-pole analysis for non-Markovian system: memory turnover}

\begin{figure}	
	\centering
	\includegraphics[width=8cm]{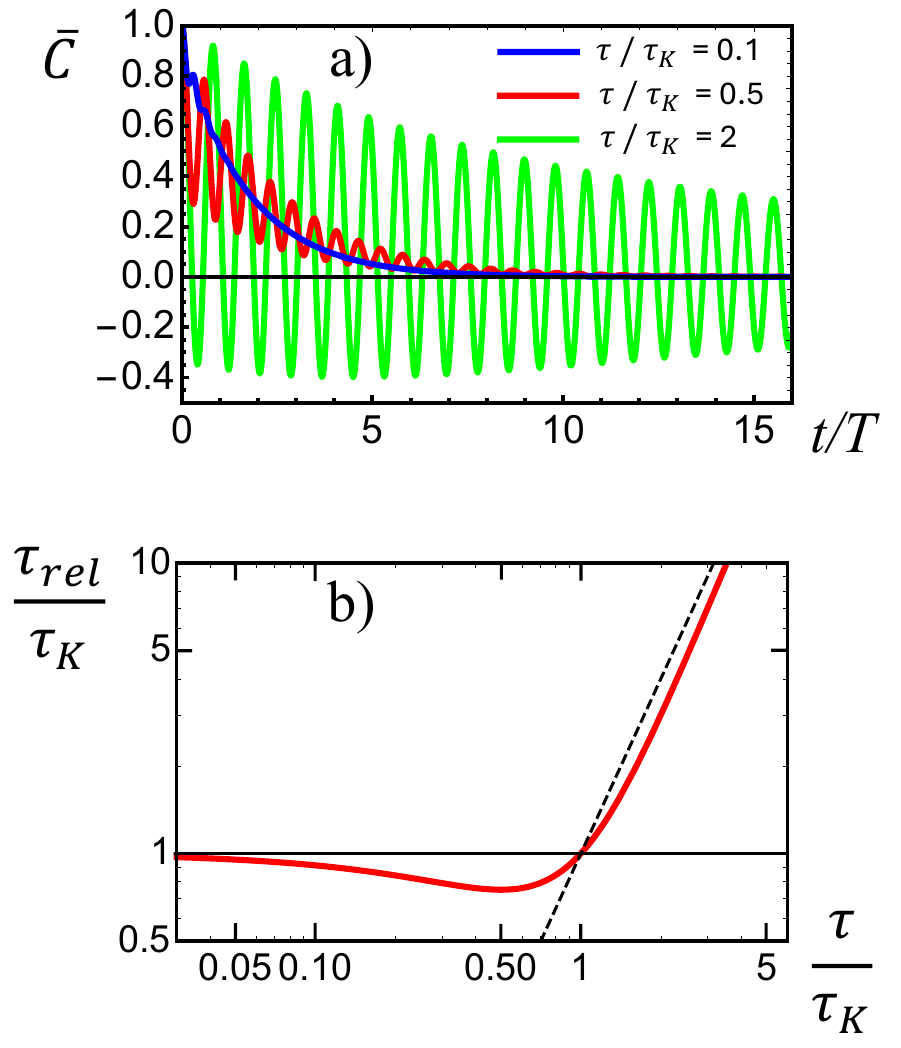}
\caption{ Analytical results for the non-Markovian three-pole scenario for finite  memory time $\tau$.
a)  Rescaled correlation function $\bar C(t)=C(t)/C_0$ according to Eqs. (\ref{GLE23}) and  (\ref{GLE24})
 using the results for the poles in Eq. (\ref{GLE22}) for three different rescaled memory times $\tau/\tau_K$,
 where $\tau_K=\gamma/K$ is the overdamped Markovian  relaxation time. 
Time $t$  is  rescaled by the Markovian harmonic oscillation period in the frictionless limit $T=2 \pi \sqrt{m/K}$
and  the rescaled mass is set to a small value of $\tau_m/\tau_K=mK/\gamma^2=0.01$. 
b)  Rescaled relaxation  time $\tau_{\rm rel} /\tau_K$ in the zero-mass limit  according to Eqs. (\ref{GLE28})  
as function of the rescaled memory time $\tau/\tau_K$. 
The asymptotic scaling in the long-memory-time  limit 
$\tau_{\rm rel} \sim \tau^2$ is indicated by a straight broken line.
		}
	\label{fig3}
\end{figure}

We now consider the response function in Eq. (\ref{GLE5}) with the single exponential 
 memory kernel in Eq. (\ref{GLE12}) for finite memory time $\tau$, 
 which we rewrite in terms of the three poles as 
\begin{align}
\label{GLE21}
 \tilde \chi(\omega)  
 &= - \frac{1+\imath \tau \omega}
  { \imath m \tau( \omega^3  - \imath  \omega^2/\tau  -\omega( \gamma/\tau +K)/m  +\imath K/(m \tau))} \nonumber \\
& = - \frac{1+\imath \tau \omega}
  { m \tau( \omega - \omega_1)(\omega - \omega_2)(\omega - \omega_3)}.
 \end{align}
Using the Cardano formula the poles are given  as 
\begin{align}
\label{GLE22}
 \omega_i
 = \epsilon_i u_+ - \frac{p}{3 \epsilon_i u_+} - \frac{b}{3},
 \end{align}
where $\epsilon_1 \equiv 1$, $\epsilon_{2,3} \equiv -1/2 \pm \imath \sqrt{3}/2$ are the complex three cubic roots of $1$, 
$u_+\equiv (-q/2+(q^2/4+p^3/27)^{1/2})^{1/3}$, $p \equiv c-b^2/3$, $q \equiv  2b^3/27 -cb/3 + d$, 
$b \equiv -\imath/\tau$, $c \equiv -\gamma/(\tau m)-K/m$, $d \equiv \imath K/(m \tau)$,
for details see Appendix  \ref{AppCardano}.
The  response function in the time domain follows  from residual calculus as
\begin{widetext}
\begin{align}
\label{GLE23a}
 \chi(t) =  -\frac{ \theta(t)  }{m\tau}  \left(
 \frac{ (1+   \imath \omega_1 \tau )  e^{\imath \omega_1 t}}{  (\omega_1-\omega_2)( \omega_1 - \omega_3)}
  + \frac{ (1+   \imath \omega_2 \tau)   e^{\imath \omega_2 t}}{  (\omega_2 - \omega_1) ( \omega_2 - \omega_3)}
   + \frac{ (1+   \imath \omega_3 \tau)   e^{\imath \omega_3 t}}{  (\omega_3 - \omega_1) ( \omega_3 - \omega_2)}
   \right),
   \end{align}
from which the  correlation function follows from Eq. (\ref{GLE10})  by integration  as
\begin{align}
\label{GLE23}
 C^+(t) =  \frac{ \theta(t) k_BT }{m\tau}  \left(
 \frac{ (\tau - \imath/\omega_1)  e^{\imath \omega_1 t}}{  (\omega_1-\omega_2)( \omega_1 - \omega_3)}
  + \frac{ (\tau - \imath/\omega_2)   e^{\imath \omega_2 t}}{  (\omega_2 - \omega_1) ( \omega_2 - \omega_3)}
   + \frac{ (\tau - \imath/\omega_3)   e^{\imath \omega_3 t}}{  (\omega_3 - \omega_1) ( \omega_3 - \omega_2)}
   \right)
   \end{align}
\end{widetext}
with the variance given  as 
\begin{align}
\label{GLE24}
 C_0=2 C^+(0) =  - \frac{ \imath k_BT}{m \tau  \omega_1 \omega_2\omega_2} = \frac{k_BT}{K}, 
   \end{align}
which again  is the expected result from the equipartition theorem. 
In Fig. \ref{fig3}a) we plot  $\bar C(t)=C(t)/C_0$
 for three different rescaled memory times $\tau/\tau_K$
as a function of time $t$  rescaled by the  oscillation period  of the Markovian  frictionless 
harmonic oscillator $T=2 \pi \sqrt{m/K}$  for fixed  small rescaled mass $\tau_m/\tau_K=mK/\gamma^2=0.01$. 
We see that for small memory time, $\tau/\tau_K=0.1$ (blue line), the correlation function decays almost purely exponentially, while for 
intermediate memory time, $\tau/\tau_K=0.5$ (red line), pronounced oscillations are visible but  the decay time of the
envelope is  not changed much.
For long memory time, $\tau/\tau_K=2$ (green line), 
we see that the oscillation period has   slightly changed 
and is now of the order of $T$, while 
the  decay time of the envelope  is increased considerably.
In Appendix  \ref{AppPole} we explain this behavior in terms of a pole analysis, where we show
that the decay time of the exponential envelope in fact diverges as $\tau^2$ for long memory times. 
These results show that for long memory time, the dynamics of the reaction coordinate becomes similar 
to a highly inertial system,
as has been noted before based on simulation trajectories 
\cite{Straub_1986,Kappler_2018, Kappler_2019, Lavacchi_2020},
 meaning that the effect of friction is effectively reduced.
The presence of oscillations with a period of $T$
 suggests that a finite system mass $m$ is crucial in order to correctly account 
for the effects of long memory times, as will be discussed in more detail in Sec.  \ref{Mass}.

From Eqs. (\ref{GLE23}) and  (\ref{GLE24}) the relaxation time defined in  Eq. (\ref{FP16}) follows
as 
\begin{widetext}
\begin{align}
\label{GLE25}
\tau_{\rm rel} & =  \frac{2K^2}{m^2 \tau^2} \left[ 
I_1( \omega_1, \omega_2,  \omega_3) +I_2( \omega_1, \omega_2,  \omega_3) 
+ I_1( \omega_2, \omega_3,  \omega_1) +I_2( \omega_2, \omega_3,  \omega_1) 
+ I_1( \omega_3, \omega_1,  \omega_2) +I_2( \omega_3, \omega_1,  \omega_2) \right],
\end{align}
\end{widetext}
where we defined the functions
\begin{align}
\label{GLE26}
I_1( \omega_1, \omega_2,  \omega_3) \equiv 
\frac{ \imath  \left(\tau - \imath/\omega_1\right)^2 }
{2  \omega_1 (\omega_1-\omega_2)^2  (\omega_1-\omega_3)^2},
\end{align}
\begin{align}
\label{GLE27}
I_2( \omega_1, \omega_2,  \omega_3)\equiv -
\frac{ 2\imath \left(\tau - \imath/\omega_2\right)  \left(\tau - \imath/\omega_3\right) }
{   (\omega_2 + \omega_3) (\omega_2-\omega_1)  (\omega_3-\omega_1) (\omega_2-\omega_3)^2}.
\end{align}
In   the zero-mass limit $m \rightarrow 0$, Eq. (\ref{GLE25})
yields
\begin{align}
\label{GLE28}
\tau_{\rm rel}  =  \frac{\gamma}{K} \left(
\frac{1+\left( \tau K /\gamma \right)^3}{1+   \tau K /\gamma} \right).
\end{align}
with the  limits given by 
\begin{align}
\tau_{\rm rel}    &\simeq    
\begin{cases}
\gamma/K - \tau   ~&,~  \tau  \ll \gamma/K    \\  
\tau^2  K/  \gamma    ~&,~  \tau    \gg \gamma/K, \end{cases} 
\label{GLE29}
 \end{align}
 see Appendix  \ref{AppPole} for details.
So we obtain for small memory times $\tau$ a reduction of  $\tau_{\rm rel}$ that is  linear
in $\tau$, which leads to the memory-induced speed-up of barrier-crossing times, while
for large memory times $\tau_{\rm rel}$ increases quadratically with $\tau$, which leads 
to  memory-induced  barrier-crossing slow-down. 
The memory-induced speed-up term we obtain in Eq. (\ref{GLE25}) for small $\tau$ is in exact agreement with the 
zero-mass prediction of GH theory, as demonstrated in Appendix  \ref{AppGH},
however, the turnover to 
 the memory-induced  barrier-crossing slow-down is  missed by GH theory.
This salient memory-turnover behavior
 is illustrated  in Fig. \ref{fig3}b, where we plot Eq. (\ref{GLE28}) as a function of the rescaled memory time
$\tau/\tau_K$. The minimum of  $\tau_{\rm rel}$ appears for memory time  $\tau^* = \gamma/(2K) $,
so we see that in the limit of a vanishing confining harmonic potential, i.e. for $K \rightarrow 0$, 
our Gaussian analysis predicts the minimum of the relaxation time to move to infinite memory time.
This might explain why GH theory, which treats the escape over a barrier without 
relaxation in a well, does not yield a minimum of the relaxation time and also not the long-memory-time 
slow-down of the barrier-crossing rate.
For memory times $\tau \gg  \tau^*$ the relaxation time exhibits 
 a quadratic increase. This increase of the relaxation time is reflected
by the slow decay of the correlation function envelope visible in Fig. \ref{fig3}a for long memory times. 
When taking the $m \rightarrow 0$ limit of the correlation function Eq. (\ref{GLE23}) before 
calculating the relaxation time according to  Eq. (\ref{FP16}) one misses the long-time barrier-crossing slow-down,
as shown in   Sec. \ref{Mass}, 
which demonstrates  that the barrier-crossing slow down for  long memory time  originates from the  
interplay between the three poles and involves a finite mass.

If instead of using $p=2$ for calculating the moments according to  Eq. (\ref{FP10}) 
one takes $p=1$, the relaxation time is instead of Eq. (\ref{GLE25}) given by 
an integral over the correlation function 
$\tau^{p=1}_{\rm rel}  =  \int_0^\infty {\rm d}t {\bar C}(t)
= \int_0^\infty {\rm d}t { C}(t)/C_0$.
From Eqs. (\ref{GLE6}) we obtain 
$\int_0^\infty {\rm d}t { C}(t) = \tilde C(\omega=0)/2=  k_BT \gamma/K^2$, 
together with $C_0=k_BT/K$ we thus obtain
$\tau^{p=1}_{\rm rel}  = \gamma/K$ independent of mass and of memory time. 
So we see that when defining the relaxation time as the integral over the correlation function, 
inertial and non-Markovian effects are completely missed, while using $p=2$ and thus
defining the relaxation as the integral over the correlation function squared, which corresponds
to the energy-energy correlation function, non-Markovian effects are accounted for.
This is easy to understand, since,
as shown in Figs. \ref{fig2}(a) and \ref{fig3}(a),
 for large mass and large memory time oscillations appear
in the correlation function and the intergral over the correlation does not reflect
the decay time of the  exponential envelope.

\subsection{ The mass-less non-Markovian system with single-exponential memory
 is singular } \label{Mass}

We reconsider the response function in Eq. (\ref{GLE5}) with the single exponential 
 memory kernel in Eq. (\ref{GLE12}) but already now take  the mass-less limit $m\rightarrow0$, 
 in which case we obtain
\begin{align}
\label{GLE30}
 \tilde \chi_{m=0}(\omega) &= \frac{1}{ K   +  \imath  \gamma \omega/(1+ \imath  \tau \omega)}=
 \frac{1 + \imath  \tau \omega}{K + \imath \omega ( \gamma + \tau K)} \nonumber \\ 
 &= \frac{- \imath \gamma}{(\gamma + \tau K)^2(  \omega - \imath K /(\gamma + \tau K)) }
 + \frac{\tau}{\gamma + K \tau},
 \end{align}
where in the last line we separated off a constant so that the remainder of the response function
decays to zero for $\omega \rightarrow \infty$. Using this separation, 
the response function in the time domain is straightforwardly obtained as 
\begin{align}
\label{GLE31}
 \chi_{m=0}(t) &=  \theta(t) \frac{\gamma}{(\gamma + \tau K)^2} e^{-t K /((\gamma + \tau K)} +  \frac{\tau}{\gamma + K \tau} \delta(t).
  \end{align}
We see that  the response function exhibits  a very unusual  delta-peak response at zero time. This singularity is
a result of neglecting the mass in the presence of a  finite memory time, 
which transpires  from the fact that 
 the delta peak vanishes for  $\tau \rightarrow 0$. As a consequence of the presence of this singularity, the 
 separation of the Fourier transform of the antisymmetric 
 position-velocity correlation function $\tilde C_{xv}(\omega)$ in  Eq. (\ref{GLE8}) into the 
 single-sided  function $\tilde C^+_{xv}(\omega)$ in  Eq. (\ref{GLE9})
 is not possible. Instead, we use  Eq. (\ref{GLE6}) to calculate  $\tilde C(\omega)$ from  Eq. (\ref{GLE31}) 
 and obtain after back Fourier transformation
\begin{align}
\label{GLE32}
 C_{m=0}(t) &=  \frac{k_BT }{K (1 + \tau K/\gamma)} e^{- |t| K /((\gamma + \tau K)}.
  \end{align}
The variance follows from this expression as $C_{m=0}(0)= k_BT/(K (1 + \tau K/\gamma))$, in stark contrast to
the equipartition-theorem prediction $C(0)= k_BT/K$ obtained from the three-pole analysis 
in Eq. (\ref{GLE24}), which is a consequence of  neglecting the 
regularizing mass in the starting equation, as has been pointed out before \cite{Nascimento2019}. 
From  Eq. (\ref{GLE32}) the relaxation time defined in  Eq. (\ref{FP16}) follows as 
$\tau^{m=0}_{rel}= (\gamma/K)(1+\tau K/\gamma)$, which predicts a linear increase of the relaxation with memory time $\tau$,
very different from the result from the full three-pole analysis in Eq. (\ref{GLE28}).
A heuristic fix is possible by rescaling the result for the relaxation time by the ratio of the mass-less and  the actual 
variances according to $ (C_{m=0}(0) /C(0))^2  \tau^{m=0}_{rel} = (\gamma/K)/(1+\tau K/\gamma)$,
which correctly recovers the intermediate memory acceleration regime  but misses the long-memory slow-down regime 
in  Eq. (\ref{GLE28}). We conclude that in order to correctly predict barrier-crossing dynamics one needs to keep a finite mass
in the equation of motion   describing the reaction-coordinate dynamics.

\subsection{The zero-memory-time limit of the  GLE is  time-reversible} \label{reversibility} 

After having shown that the zero-mass limit of the GLE is singular, 
we now discuss the zero-memory-time limit $\tau \rightarrow 0$
of the GLE and show that in this limit also a singularity arises, which 
in fact is crucial in order to preserve time reversibility of the resulting Langevin equation. 
This resolves the puzzling finding that 
in its usual formulation,   the Langevin equation breaks time-reversal symmetry, 
while the GLE,  which is derived by  exact projection  from 
Hamilton's equation of motion, obeys time-reversal symmetry (as Hamilton dynamics does).

To clarify this issue, we consider the Mori GLE in the form obtained by  projection formalism 
\cite{Mori_1965, zwanzig_nonequilibrium_2001}, which reads \cite{Netz2025c}
\begin{widetext}
\begin{align} \label{rev1}
  \ddot B(w, t)  = -  K_M  (B(w,t) - \langle B \rangle ) 
   - \int_0^{t-t_P} {\rm d}s\, \Gamma_M(s) \dot B(w, t-s) + F_M(w,t).
\end{align}
\end{widetext}

Note that in contrast to the GLE Eq. \eqref{GLE1} we have been using to derive all explicit results so far, 
the Heisenberg observable $B(w, t)$ and the orthogonal force $F_M(w,t)$  in  Eq.~\eqref{rev1} have an explicit dependence
on the $6N$-dimensional phase-space variable $w=( \{{\bf r}_N\} , \{{\bf p}_N\} ) $, where $N$ denotes the number of particles
and $ \{{\bf r}_N\}$ are the particle positions and $ \{{\bf p}_N\}$  the particle momenta in three-dimensional space.
Time reversibility means
that the  trajectory of a many-body system is invariant when time and momenta are inverted. The Hamilton equations
of motion are time-reversible for a Hamiltonian that is even in  the momenta $\{{\bf p}_N\}$.
Note that   eq.~\eqref{rev1} results from an  exact derivation and therefore must also be  time-reversible.
 The time at which the projection is taking place, $t_P$,  appears explicitly in Eq. \eqref{rev1} \cite{Netz2025c}.
 Eq. \eqref{GLE1} is obtained from Eq. \eqref{rev1} by i) pushing the propagation time   in Eq. \eqref{rev1} 
infinitely far into the past, $t_P \rightarrow  - \infty$,
which produces the time ordering $t>t_P$ and  makes the upper integral boundary   in Eq. \eqref{rev1}  equal to 
$+\infty$, ii) defining mass-rescaled parameters $\Gamma(t) =m \theta(t) \Gamma_M(t)$, $K=m K_m$ and 
$F(t)= m F_M(w,t)$,  where we defined the memory kernel $\Gamma(t)$ to be single-sided so that the 
integration range of the friction term can be extended to the negative time domain,
iii) introducing a shifted observable $x(t) = B(w, t)- \langle B \rangle$ and skipping  the phase-space
dependence of x(t) and F(t).

Setting $t_P=0$ in the remainder, time reversibility  around $t=0$ manifests for the Heisenberg observable as 
\begin{align} \label{rev2}
  \ddot B(\{{\bf r}_N\},\{{\bf p}_N\},  t) & =  \ddot B(\{{\bf r}_N\},- \{{\bf p}_N\},-t) \nonumber \\
    \dot B(\{{\bf r}_N\},\{{\bf p}_N\}, t) & = - \dot B(\{{\bf r}_N\}, - \{{\bf p}_N\}, -t)\nonumber \\
        B( \{{\bf r}_N\},\{{\bf p}_N\} , t) & =  B(\{{\bf r}_N\}, - \{{\bf p}_N\}, -t).
  \end{align}
Using the time-reversal symmetry of $\Gamma_M(t)$, i.e. 
$\Gamma_M(t) = \Gamma_M(-t)$,
we obtain 
\begin{align} \label{rev3}
 &  \int_0^{t} {\rm d}s\, \Gamma_M(s) \dot B(\{{\bf r}_N\},\{{\bf p}_N\} , t-s) \nonumber \\ 
  & =  \int_0^{-t} {\rm d}s\, \Gamma_M(s) \dot B( \{{\bf r}_N\},-\{{\bf p}_N\}, -t-s),
\end{align}
thus the memory-friction term satisfies time-reversal symmetry as well. 
Since  the left side and the first two terms of the right side of Eq. \eqref{rev1}
are time-reversible, realizing that $K_M$ is independent of time and momenta, 
 the orthogonal force must also be time-reversible  since the entire equation
is, we therefore conclude 
\begin{align} \label{rev4}
 F( \{{\bf r}_N\},\{{\bf p}_N\}, t) =
 F( \{{\bf r}_N\},- \{{\bf p}_N\}, -t).
 \end{align}
Thus not only the GLE as an equation is time-reversible, in fact each individual term of the GLE is time-reversible as well, 
which  is a much stronger statement. 
For details and derivations see Appendix \ref{AppMori}, 

In the Markovian limit the memory kernel becomes infinitely short-ranged and can be written as 
\begin{align} \label{rev5}
 \Gamma_M(t) = 2 \gamma_M \delta(t).
  \end{align}
Using that 
\begin{align} \label{rev6}
& \int_0^{t} {\rm d}s\, \Gamma_M(s) \dot B(w, t-s) \nonumber \\
&= 
2 \gamma_M   \int_0^{t} {\rm d}s\, \delta(s) \dot B(w, t-s) \nonumber \\
&=  2 \gamma_M \dot B(w, t)   \int_0^{t} {\rm d}s\, \delta(s) =
 \gamma_M \dot B(w, t)   {\rm sig}(t),
  \end{align}
we obtain in  the Markovian limit the  Langevin equation as
\begin{align} \label{rev7}
  \ddot B(w, t)  =& -  K_M  (B(w,t) - \langle B \rangle ) 
   - \gamma_M  {\rm sig}(t)  \dot B(w, t) \nonumber \\ 
 &  + F_M(w,t),
\end{align}
where $ {\rm sig}(t)$ is the signum function with $ {\rm sig}(t)=1$ for $t>0$ and  $ {\rm sig}(t)=-1$ for $t<0$.
Obviously, the Langevin equation in Eq. \eqref{rev7}, obeys time-reversal symmetry around $t=0$ 
but exhibits a discontinuity
of the friction term  at the time origin, $t=0$, which 
reflects the fact that the friction term in the underlying GLE in Eq. \eqref{rev1} vanishes for $t-t_P=0$. 
We note that this discontinuity of the friction term in the Langevin Eq. \eqref{rev7} is unproblematic
when considering time-averaged quantities. We therefore  not only show in Sec. \ref{Mass}
 that the zero-mass limit is ill-defined for non-Markovian friction kernels (at least for the single-exponential
 kernel  we consider in this paper), 
 we also show here  that in the zero-memory-time limit  the resulting Langevin equation is time-reversible 
 and contains a discontinuity 
 which, when neglected, breaks time-reversal symmetry.

\begin{figure}	
	\centering
	\includegraphics[width=8cm]{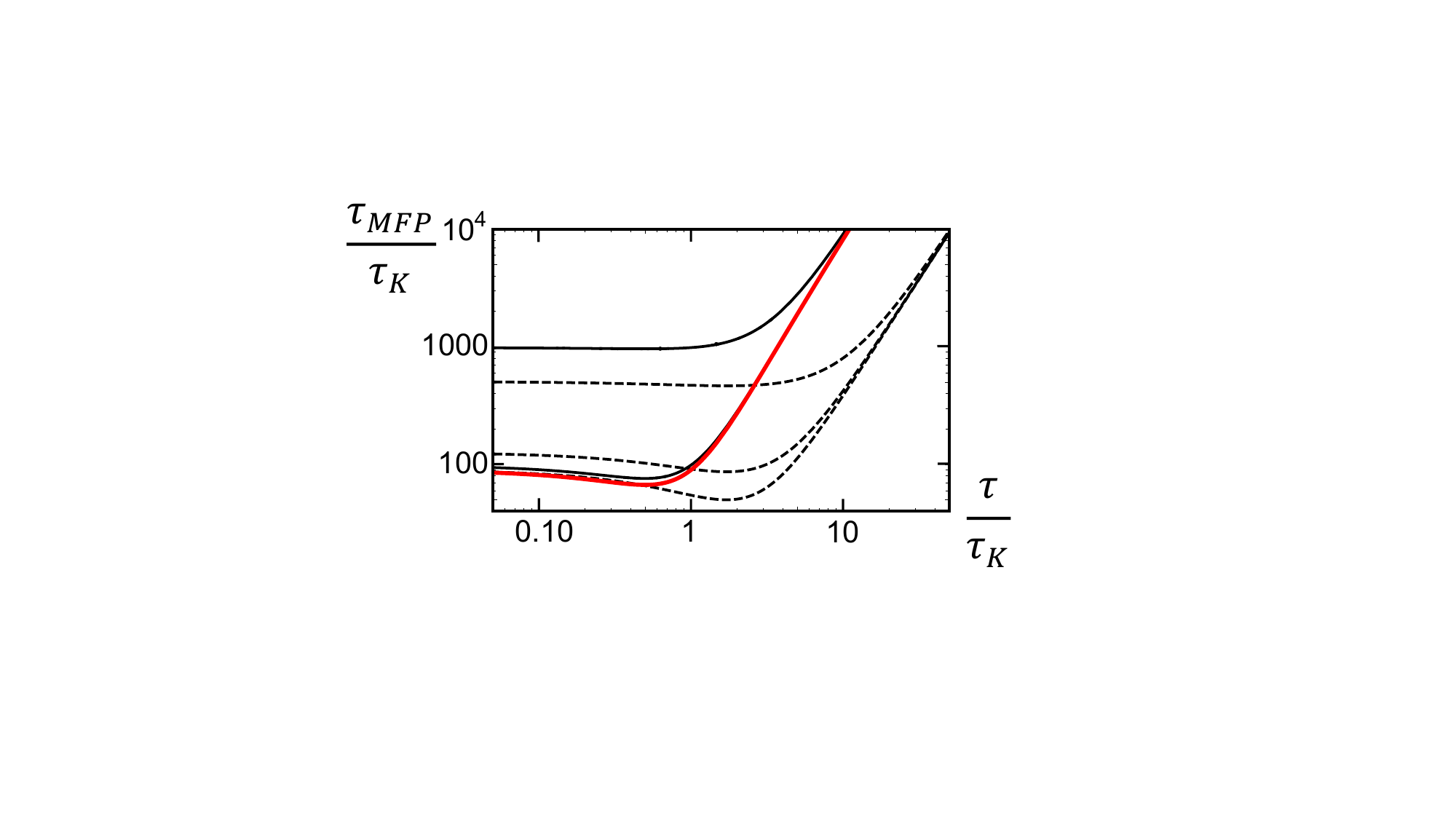}
\caption{  Comparison of our 
analytical results 
(Eqs. \eqref{final3}, \eqref{final4} and  \eqref{GLE25},  black solid lines, 
Eqs. \eqref{final3}, \eqref{final4} and  \eqref{GLE28},  red solid line) 
and a previous fit of a scaling function  to simulation results (Eq. 8 in  \cite{Lavacchi2025}, broken lines)
for the well-to-barrier-top mean-first passage time (rescaled by $\tau_K=\gamma/K$) 
  of   the non-Markovian model with single-exponential memory, all for a barrier height of $U_0 = 3 k_BT$.
 Results are shown for  three different values of the  rescaled mass
  $\tau_m/\tau_K=mK/\gamma^2=10, 0.1, 0$ (from top to bottom)
  as a function of the memory time rescaled by $\tau_K=\gamma/K$.
 		}
	\label{fig4}
\end{figure}

\section{  Conclusions and Discussion} \label{Conclu}

Using an exact  mapping of the Green function  of a  Gaussian general non-Markovian reaction coordinate $x$
onto a generalized Fokker Planck equation with a time-dependent diffusivity, 
we derive an expression of the mean-first passage to reach $x_f$ starting from $x_0< x_f$
which, combining  Eqs. \eqref{MFP12},  (\ref{FP15}), (\ref{FP16}), reads 
\begin{align}
\label{final1}
\tau_{\rm mfp}  = \frac{ \tau_{\rm rel}  }{C_0}   \int_{x_0}^{x_f} {\rm d}x e^{ \beta U(x) }
 \int_{x_{\rm ref}}^{x} {\rm d}x' e^{ -  \beta U(x') }.
 \end{align}
In this expression $x_{\rm ref}< x_0$ is the position of a reflecting boundary, 
 $C_0=C(t=0)=\langle x^2 \rangle $ is the mean-square position of the reaction coordinate and
 $\tau_{\rm rel}$ denotes the relaxation time given by an integral over the squared
 positional correlation function $C(t)=\langle x(0) x(t) \rangle $ according to 
\begin{align}
\label{final2}
\tau_{\rm rel} & 
=\frac{2}{C_0^2}  \int_0^\infty {\rm d}t { C}^2(t).
\end{align}
The derivation is exact  for a harmonic potential $U(x)=Kx^2/2$ and
for a correlation function $C(t)$ that decays single-exponentially, 
for general $C(t)$ Eq. (\ref{final2}) correspond to the first term of  a systematic cumulant expansion. 
Employing a saddle point approximation on the double-integral in Eq. (\ref{final1})  we obtain
the Arrhenius-type expression in terms of the exponential of the barrier height $U_0$
\begin{align}
\label{final3}
\tau_{\rm mfp}  = \tau_{\rm rel} \Theta   e^{ \beta U_0 },
 \end{align}
where $\Theta$ is a unitless factor that depends on the potential shape. 
For the three  scenarios shown in Fig. \ref{fig1} it is given by  
(Eqs. (\ref{pot1}), (\ref{pot2}), (\ref{pot3}))
\begin{align}
\Theta  &= 
\begin{cases}
 \sqrt{  \frac{ \pi}{\beta U_0 } }      ~&~  {\rm scenario \,\,\,  Fig. 1a}     \\  
\pi    \left( \frac{K }{ K_{\rm bar}}\right)^{1/2}       ~&~  {\rm scenario \,\,\, Fig. 1b}  \\
2\pi    \left( \frac{K }{ K_{\rm bar}}\right)^{1/2}       ~&~  {\rm scenario \,\,\,  Fig. 1c} .
\end{cases} \label{final4}
 \end{align}
One key feature  of our theory, which distinguishes it from other theories,  is that memory and inertial
effects are fully included and enter 
via the two-point auto-correlation function $C(t)$ , which  from simulations and  experiments can
be much more easily 
determined than the memory function. 
In Fig. \ref{fig2}b we demonstrate that the so-called Kramers turnover, i.e. the minimum of 
the barrier-crossing time of a Markovian system  for intermediate friction and the  asymptotic scaling 
 in the high and low friction limits,
$\tau_{\rm mfp} \sim (\gamma/K)  e^{ \beta U_0 }$ and $\tau_{\rm mfp} \sim (m/ \gamma) e^{ \beta U_0 }$
 are correctly reproduced. In  Fig. \ref{fig3}b we demonstrate
that also the more complex memory turnover, i.e.  the minimum of the barrier-crossing time for intermediate 
memory time $\tau$ and the asymptotic scaling $\tau_{\rm mfp} \sim \tau^2 e^{ \beta U_0 }$,  
are correctly reproduced. 
It is  noteworthy that our theory reproduces exactly  the GH theory in the zero-mass small-memory-time
limit but, in contrast to GH theory, also correctly reproduces the large-memory-time limit.

In Fig. \ref{fig4} we perform a quantitative comparison
 for the mean-first passage time to reach from the well minimum to the barrier top (scenario Fig. \ref{fig1}b)
 predicted by  our theory
 with   extensive simulation data for a double-well potential and a single-exponential kernel  \cite{Lavacchi2025}.
The mean-first passage  times in the simulations are obtained from  long
trajectories by calculating the mean all-to-first passage times, i.e. counting
all recrossings of the starting position, which has been shown to correspond to
the escape time from a well, which is the quantity predicted by reaction rate theories \cite{Zhou_2025}.
We compare our theory 
(Eqs. \eqref{final3}, second line of  \eqref{final4} and  \eqref{GLE25},  black solid lines, 
Eqs. \eqref{final3}, second line of  \eqref{final4} and  \eqref{GLE28},  red solid line) 
for three different values of the  rescaled mass
  $\tau_m/\tau_K=mK/\gamma^2=10, 0.1, 0$ (from top to bottom)
with a  scaling function calibrated  to simulation results 
(Eq. 8 in  \cite{Lavacchi2025}, broken lines). The simulations were done for a double well potential of the form 
$U(x)= U_0(x^2/L^2-1)^2$, for which $K/K_{\rm bar}=2$, we use $U_0=3 k_BT$ in Fig. \ref{fig4}
since the simulations were mostly done around that  barrier energy.
We see that our theory and the simulation data agree well for low memory time and low mass, 
which reflects that in this limit our theory becomes exact. For large mass and long memory times
there are deviations,  yet, our theory reproduces the memory turnover,
that means the minimum in $\tau_{\rm mfp}$ as a function of $\tau$, rather well. 

The deviations between our theoretical predictions (solid lines) and the previously
published simulation results (broken lines) in Fig. \ref{fig4} stem from two approximations we employed.
First, our derivation is exact for a harmonic potential and treats the barrier as an absorbing cusp,
as illustrated in Fig. \ref{fig1}a. We obtain on the harmonic level a separation between potential and
memory/inertial effects, as seen in Eqs. \eqref{final1}, for non-harmonic potentials we expect  a 
coupling beween non-harmonicities and memory effects, which could be studied in simulations.
Secondly, our derivation is exact for single-exponential correlation functions $C(t)$
and treats more complex correlation functions by a systematic cumulant expansion 
to first order (see Sec. \ref{Cumulant}).
As Fig. \ref{fig3}b demonstrates, for long memory time $\tau$ the correlation function
even for single-exponential memory kernels  does not decay single-exponentially
but exhibits an oscillating-exponential decay. In future extensions of our theory, it would 
be interesting to go beyond the first-order cumulant expansion.

\begin{acknowledgments}
We acknowledge support by Deutsche Forschungsgemeinschaft Grant CRC 1449 
"Dynamic Hydrogels at Biointerfaces ", 
Project ID 431232613, Projects A02 and A03.
\end{acknowledgments}

\appendix

\section{Derivation of the Green function \label{AppGreen}}

The condition in Eq. \eqref{dist2} that     the joint distribution is symmetric  leads to $c(t)=a(t)$. 
The normalization in Eq. \eqref{dist3}  leads to 
\begin{align}
\label{AppGreen1}
{\cal N}= \frac{2 \pi}{\sqrt{a^2(t)-b^2(t)}}.
\end{align}
The definition of the two-point autocorrelation function  $C(t)$ in Eq. \eqref{dist4}  leads to 
\begin{align}
\label{AppGreen2}
C(t)= -\frac{b(t)}{a^2(t)-b^2(t)}.
\end{align}
The definition of the variance $C_0$  in Eq. \eqref{dist5}  leads to 
\begin{align}
\label{AppGreen3}
C_0= \frac{a(t)}{a^2(t)-b^2(t)}.
\end{align}
Inversion of these equations leads to 
\begin{align}
\label{AppGreen4}
a(t)= \frac{1}{C_0 \left( 1-\bar C^2(t) \right) }
\end{align}
and
\begin{align}
\label{AppGreen5}
b(t)= -  \frac{\bar C(t)}{C_0 \left( 1-\bar C^2(t) \right) },
\end{align}
where we used the rescaled correlation function $\bar C(t)$  defined in Eq. \eqref{dist7} 
and from which Eq. \eqref{dist6}  follows.

\begin{widetext}
\section{Derivation of the generalized Fokker-Planck equation   \label{AppFP}}

By taking a time derivative of Eq. \eqref{dist9} we obtain

\begin{align}
\label{AppFP1}
\frac{ \partial {\cal P}(x,t | x_0) }{\partial t}&=
\frac{\bar C'(t)
 \left(
\left(x- \bar C(t)x_0 \right) x_0  \bar C(t) \left(1- \bar C^2(t) \right)
- \left(x- \bar C(t)x_0 \right)^2 \bar C^2(t)
+  \left(1- \bar C^2(t) \right) C_0 \bar C^2(t) \right)}
{C_0 \bar C(t) \left(1- \bar C^2(t) \right)^2}
 {\cal P}(x,t | x_0),
\end{align}
where $\bar C'(t)$ denotes the derivate of $\bar C(t)$.
With the first and second spatial derivatives of Eq. \eqref{dist9},
\begin{align}
\label{AppFP2}
\frac{ \partial {\cal P}(x,t | x_0) }{\partial x}&=-
\frac{ \left(x- \bar C(t)x_0 \right)}
{C_0 \left(1- \bar C^2(t) \right)}
{\cal P}(x,t | x_0),
\end{align}
\begin{align}
\label{AppFP3}
\frac{ \partial^2 {\cal P}(x,t | x_0) }{\partial x^2}&=
\left(- \frac{ 1 } {C_0 \left(1- \bar C^2(t) \right)} +
\frac{ \left(x- \bar C(t)x_0 \right)^2}
{C_0^2 \left(1- \bar C^2(t) \right)^2} \right)
{\cal P}(x,t | x_0),
\end{align}
we obtain by comparison with Eq. \eqref{AppFP1}
\begin{align}
\label{AppFP4}
\frac{ \partial {\cal P}(x,t | x_0) }{\partial t}&=
- \frac{\bar C'(t)}{\bar C(t)} {\cal P}(x,t | x_0)
- x  \frac{\bar C'(t)}{\bar C(t)} \frac{ \partial {\cal P}(x,t | x_0) }{\partial x}
- \frac{C_0 \bar C'(t)}{\bar C(t)} \frac{ \partial^2 {\cal P}(x,t | x_0) }{\partial x^2}
\end{align}
which is equivalent to  Eq. \eqref{FP1}.

\end{widetext}

\section{Saddle-point approximation of the 
mean-first passage time for different potential shapes    \label{AppMFPT}}

To first order in the cumulant expansion introduced in Section \ref{Cumulant},
the mean-first passage time  to
reach the final position $x_f$ for the first time when  starting from  $x_0$ follows from 
 Eq. \eqref{MFP12} as
\begin{align}
\label{AppSP1}
 \tau_{\rm mfp}  = D_0^{-1}   \int_{x_0}^{x_f} {\rm d}x e^{ \beta U(x) }
 \int_{x_{\rm ref}}^{x} {\rm d}x' e^{ -  \beta U(x') }.
 \end{align}
For high barriers, this double integral can be estimated using  the saddle-point approximation. 
Let us first consider the  scenario depicted in Fig. \ref{fig1}c, where the initial position  $x_0$
is located to the left of the barrier and the  final position $x_f$ is located to the right of the barrier. 
We write 
\begin{align}
\label{AppSP2}
 \tau_{\rm mfp}  = D_0^{-1}   \int_{x_0}^{x_f} {\rm d}x e^{ \beta U(x) } I(x).
 \end{align}
 Since the integral over $x$ includes the barrier,
the exponential function in the integrand is maximal at the barrier position $x=x_{\rm max}$, we therefore expand
the  function $I(x)$ around $x=x_{\rm max}$ and obtain to first order
\begin{align}
\label{AppSP3}
 I(x) & \equiv  \int_{x_{\rm ref}}^{x} {\rm d}x' e^{ -  \beta U(x') } \nonumber  \\ 
 &= I(x_{\rm max}) + (x-x_{\rm max})  I'(x_{\rm max}) + \cdots \nonumber  \\ 
 & =  \int_{x_{\rm ref}}^{x_{\rm max}} {\rm d}x' e^{ -  \beta U(x') } + (x-x_{\rm max}) e^{ -  \beta U(x_{\rm max} )}  + \cdots
 \end{align}
The integrand is maximal at the minimum in the left well $x_{\rm min}$, which is included in the integral 
for  $x_{\rm ref} < x_{\rm min} < x_{\rm max}$. Assuming a high barrier and correspondingly a deep minimum
in the well, the integral boundaries can be pushed
to infinity and the potential can be expanded to second order around the minimum,  so that we obtain 
\begin{align}
\label{AppSP4}
 I(x)  \simeq &  \int_{-\infty}^{\infty} {\rm d}x' e^{ -  \beta U(x') } + 
 (x-x_{\rm max}) e^{ -  \beta U(x_{\rm max} )}  + \cdots \nonumber  \\ 
  \simeq &  \int_{-\infty}^{\infty} {\rm d}x' e^{ -  \beta U(x_{\rm min}) -\beta K (x'-x_{\rm min})^2/2 } \nonumber  \\ 
& + (x-x_{\rm max}) e^{ -  \beta U(x_{\rm max} )}  + \cdots \nonumber  \\ 
  \simeq &  e^{ -  \beta U(x_{\rm min})} \sqrt{\frac{2 \pi}{\beta K }} ,  \\ 
 \end{align}
where we neglected the term proportional to $e^{ -  \beta U(x_{\rm max})}$ since for large
barriers it is negligible compared to the term proportional to $e^{ -  \beta U(x_{\rm min})}$.
To leading order $I(x)$ thus is a constant and we obtain for  Eq. \eqref{AppSP2}  
\begin{align}
\label{AppSP5}
 \tau_{\rm mfp}  \simeq  D_0^{-1}  e^{ -  \beta U(x_{\rm min})} \sqrt{\frac{2 \pi}{\beta K }} 
  \int_{x_0}^{x_f} {\rm d}x e^{ \beta U(x) }.
 \end{align}
We now expand the argument of the exponential in the 
 integrand around its maximum at $x_{\rm max}$ and extend the integration 
boundaries to plus and minus infinity, after which we obtain 
\begin{align}
\label{AppSP6}
 \tau_{\rm mfp}  
  &  \simeq  D_0^{-1}  e^{ -  \beta U(x_{\rm min})} \sqrt{\frac{2 \pi}{\beta K }} \nonumber  \\ 
&\,\, \,\, \,\,  \times   \int_{-\infty}^{\infty} {\rm d}x e^{ \beta U(x_{\rm max}) 
  -\beta K_{\rm _{\rm bar}} (x-x_{\rm max})^2/2 } \nonumber  \\ 
  &  \simeq  \frac{2 \pi}{\beta D_0 \sqrt{K K_{\rm  bar}}}   e^{   \beta U_0},
 \end{align}
where $K_{\rm  bar}$ denotes the absolute value of the curvature at the barrier
and $U_0 \equiv  U(x_{\rm max})  - U(x_{\rm min})$ is the barrier height.
Using $D_0= C_0/\tau_{\rm rel}$ and  $C_0=k_BT / K$ this is equivalent to Eq. \eqref{pot3}.

We now consider the  scenario  in Fig. \ref{fig1}b, where the initial position  $x_0$
is located to the left of the barrier and the  final position $x_f=x_{\rm max}$ is located  right at the barrier. 
We  rewrite Eq. \eqref{AppSP1}  as
\begin{align}
\label{AppSP7}
 \tau_{\rm mfp}  = D_0^{-1}   \int_{x_0}^{x_{\rm max}} {\rm d}x e^{ \beta U(x) } I(x).
 \end{align}
To leading order $I(x)$ is again given by Eq. \eqref{AppSP4}  
so we obtain for  Eq. \eqref{AppSP2}  
\begin{align}
\label{AppSP8}
 \tau_{\rm mfp}  \simeq  D_0^{-1}  e^{ -  \beta U(x_{\rm min})} \sqrt{\frac{2 \pi}{\beta K }} 
  \int_{x_0}^{x_{\rm max}} {\rm d}x e^{ \beta U(x) }.
 \end{align}
We again expand the argument of the exponential in the 
 integrand around its maximum at $x_{\rm max}$ and extend the
lower  integration 
boundary to minus infinity, after which we obtain 
\begin{align}
\label{AppSP9}
 \tau_{\rm mfp}  
  &  \simeq  D_0^{-1}  e^{ -  \beta U(x_{\rm min})} \sqrt{\frac{2 \pi}{\beta K }} \nonumber  \\ 
&\,\, \,\, \,\,  \times   \int_{-\infty}^{x_{\rm max}} {\rm d}x e^{ \beta U(x_{\rm max}) 
  -\beta K_{\rm _{\rm bar}} (x-x_{\rm max})^2/2 } \nonumber  \\ 
  &  \simeq  \frac{ \pi}{\beta D_0 \sqrt{K K_{\rm  bar}}}   e^{  \beta U_0},
 \end{align}
which using $D_0= C_0/\tau_{\rm rel}$ and  $C_0=k_BT / K$  is equivalent to Eq. \eqref{pot2}.

We now consider the  harmonic scenario  in Fig. \ref{fig1}a, where both  initial position  $x_0$
and   final position $x_f > x_0 $ are located in a harmonic potential well $U(x)=Kx^2/2$.
We  rewrite Eq. \eqref{AppSP1}  as
\begin{align}
\label{AppSP10}
 \tau_{\rm mfp}  = D_0^{-1}   \int_{x_0}^{x_f} {\rm d}x e^{ \beta U(x) } I(x).
 \end{align}
To leading order $I(x)$ is again given by Eq. \eqref{AppSP4}  
so we obtain for  Eq. \eqref{AppSP2}  
\begin{align}
\label{AppSP11}
 \tau_{\rm mfp}  \simeq  D_0^{-1}   \sqrt{\frac{2 \pi}{\beta K }} 
  \int_{x_0}^{x_f} {\rm d}x e^{ \beta U(x) },
 \end{align}
where we used that  $U(x_{\rm min})=0$. 
We  expand the argument of the exponential in the
 integrand to first order around  $x_f$ and extend the lower  integration 
boundary to minus infinity, after which we obtain 
\begin{align}
\label{AppSP12}
 \tau_{\rm mfp}  
  &  \simeq  D_0^{-1}   \sqrt{\frac{2 \pi}{\beta K }}  \int_{-\infty}^{x_f} {\rm d}x e^{ \beta U(x_f) 
  +\beta (x-x_f) U'(x_f)} \nonumber  \\ 
  &  \simeq \frac{1}{D_0 \beta  U'(x_f)}  \sqrt{\frac{2 \pi}{ \beta K }}    e^{   \beta U_0}
   \simeq \frac{1}{D_0 \beta  K }  \sqrt{\frac{ \pi}{ \beta U_0 }}    e^{   \beta U_0}
 \end{align}
where we used $U(x_f) = U_0$, $U'(x_f)= K x_f = \sqrt{2 U_0 K}$ and
which is, using $D_0= C_0/\tau_{\rm rel}$ and  $C_0=k_BT / K$, equivalent to Eq. \eqref{pot1}.

\section{ Three-pole analysis using  Cardano equation \label{AppCardano}}

 The three poles are determined from Eq. (\ref{GLE21}) by the equation
\begin{align}
\label{Card1}
  \omega^3  +b  \omega^2  + c \omega  +d=0
 \end{align}
with $b \equiv -\imath/\tau$, $c \equiv -\gamma/(\tau m)-K/m$, $d \equiv \imath K/(m \tau)$.
Inserting the definition $\omega= t-b/3$  into Eq. (\ref{Card1})  leads  to 
\begin{align}
\label{Card2}
  t^3   + p t  +q=0
 \end{align}
with $p \equiv c-b^2/3$, $q \equiv  2b^3/27 -cb/3 + d$.
Inserting now  the definition $t= u-p/(3u)$,  Eq. (\ref{Card2})  reduces to 
\begin{align}
\label{Card3}
  u^6   + q u^3  -p^3/27=0,
 \end{align}
which is a quadratic equation for $u^3$ with the two solutions
\begin{align}
\label{Card4}
  u^3_\pm = -\frac{q}{2} \pm  \sqrt{ \frac{q^2}{4}+ \frac{p^3}{27} }.
 \end{align}
Taking the cubic root we obtain a total of six possible  solutions
\begin{align}
\label{Card5}
  u_{i \pm}  =  \epsilon_i u_\pm =  \epsilon_i \left[-\frac{q}{2} \pm  \sqrt{ \frac{q^2}{4}+ \frac{p^3}{27} }\right]^{1/3},
 \end{align}
where $\epsilon_1 \equiv 1$, $\epsilon_{2,3} \equiv -1/2 \pm \imath \sqrt{3}/2$ are the complex  three cubic roots of $1$.
To  show that there are actually only three distinct solutions of  Eq. (\ref{Card3}),  we  first realize that 
\begin{align}
\label{Card6}
  u_+^3  =  - \frac{p^3}{27 u_-^3   },
 \end{align}
from which follows that 
\begin{align}
\label{Card7}
  u_+  =  - \epsilon_j \frac{p}{3 u_-   }.
 \end{align}
From the definition  $t= u-p/(3u)$  we obtain from the positive-root solutions  $\epsilon_i u_+$ 
of Eq. (\ref{Card3})  three solutions of Eq. (\ref{Card2}) as
\begin{align}
\label{Card8}
  t_i  =   \epsilon_i u_+ - \frac{p}{3 \epsilon_i u_+   }.
 \end{align}
Inserting Eq. (\ref{Card7})  into Eq. (\ref{Card8})   we obtain
\begin{align}
\label{Card9}
  t_i  =  -  \epsilon_i  \epsilon_j \frac{p}{3 u_-} + \frac{u_-}{ \epsilon_i \epsilon_j   }
 \end{align}
where $j=1,2,3$ is arbitrary. 
Using the relations 
\begin{align}
\label{Card10}
  \epsilon_1  \epsilon_1 & =  \epsilon_1 = 1/  \epsilon_1 \nonumber \\
  \epsilon_2  \epsilon_2 & =  \epsilon_3 = 1/  \epsilon_2 \nonumber \\
  \epsilon_3  \epsilon_3 & =  \epsilon_2 = 1/  \epsilon_3 \nonumber \\
 \epsilon_1  \epsilon_2 & =  \epsilon_2 = 1/  \epsilon_3 \nonumber \\
  \epsilon_1  \epsilon_3 & =  \epsilon_3 = 1/  \epsilon_2 \nonumber \\
  \epsilon_2  \epsilon_3 & =  \epsilon_1 = 1/  \epsilon_1,
   \end{align}
it immediately follows from comparing Eq. (\ref{Card8}) and Eq. (\ref{Card9}) that  
\begin{align}
\label{Card11}
  t_i  =   \epsilon_i u_+ - \frac{p}{3 \epsilon_i u_+   } = \epsilon_k u_- - \frac{p}{3 \epsilon_k u_-   }
 \end{align}
where $k=1,2,3$ is determined from the values of $i$ and $j$ according to the relations in Eq. (\ref{Card10}).
We conclude from Eq. (\ref{Card11}) that there are only three distinct solutions of  Eq. (\ref{Card2}) which can,
without loss of generality, 
be expressed using the positive root  $ u_+$ given in Eq. (\ref{Card8}).
From  the definition $\omega= t-b/3$  we finally obtain for
 the poles 
\begin{align}
\label{Card12}
 \omega_i
 = \epsilon_i u_+ - \frac{p}{3 \epsilon_i u_+} - \frac{b}{3}
 \end{align}
with $i=1,2,3$.

\section{ Asymptotic zero-mass  analysis of poles and relaxation time \label{AppPole}}

From the definitions in Appendix \ref{AppCardano} we obtain for the variables $p, q$   appearing in Eq. (\ref{Card12})
\begin{align}
\label{Pole1}
 p&= c-\frac{b^2}{3}= -\frac{\gamma/\tau +K}{m}+\frac{1}{3 \tau^2}, \\
 q&= \frac{2 b^3}{27}-\frac{cb}{3}+d
 = \frac{2 \imath}{27\tau^3} +\frac{2 \imath K}{3 m\tau}-\frac{\imath \gamma}{3 m \tau^2}.
 \end{align}
We see that both $p$ and $q$ for small mass $m$ diverge as $m^{-1}$, so the  terms involving $u_+$
in Eq. (\ref{Card12}) are to leading orders  in $m$ (including the non-divergent $m^0$ contribution) given as
\begin{align}
\label{Pole2}
u_+ \simeq \frac{p^{1/2}}{\sqrt{3}}- \frac{q}{2p} , \\
\frac{p}{3 u_+}   \simeq \frac{p^{1/2}}{\sqrt{3}}+ \frac{q}{2p} .
 \end{align}
For the poles in Eq. (\ref{Card12})  we obtain to leading orders (including the $m^0$ contribution)
\begin{align}
\label{Pole3}
\omega_1 &\simeq -\frac{q}{p} - \frac{b}{3} \simeq -\frac{\imath K}{mp \tau} \simeq \frac{\imath K}{\gamma + \tau K},\\
\omega_2 &\simeq \imath p^{1/2}+\frac{q}{2p} -\frac{b}{3} \nonumber \\
&\simeq-\left(\frac{K}{m}\right)^{1/2} \left(1+ \frac{\gamma}{\tau K}\right)^{1/2} +
\frac{\imath \gamma/\tau^2}{2K(1+\gamma/(\tau K))}, \\
\omega_3 &\simeq - \imath p^{1/2}+\frac{q}{2p} -\frac{b}{3}\nonumber \\
&\simeq\left(\frac{K}{m}\right)^{1/2} \left(1+ \frac{\gamma}{\tau K}\right)^{1/2} +
\frac{\imath \gamma/\tau^2}{2K(1+\gamma/(\tau K))}.
 \end{align}
We see that $\omega_1 \sim m^0$ while $\omega_2 \sim \omega_3 \sim m^{-1/2}$
for $m\rightarrow 0$. 
In particular, we obtain  that that $\omega_2$ and $\omega_3$ have a real part 
given by $(K/m)^{1/2}$ for large memory time $\tau$, which is the oscillation frequency
of the undamped Markovian harmonic oscillator, and an imaginary part given by
$\imath \gamma /(2 K \tau^2)$ for large memory time $\tau$, 
which  in the correlation function in Eqs. (\ref{GLE23})  leads  to a long exponential 
decay time  proportional to $\tau^2$, as indeed seen in Fig. \ref{fig3}(a).
All sums and differences of two different poles diverge as $m^{-1/2}$ except 
\begin{align}
\label{Pole4}
\omega_2 + \omega_3 \simeq q/p - 2b/ 3 \simeq \frac{\imath \gamma/\tau^2}{K(1+\gamma/(\tau K))}.
 \end{align}
This means that from the six terms contributing to $\tau_{\rm rel} $ in   Eq. (\ref{GLE25})
only the terms $I_1( \omega_1, \omega_2,  \omega_3)$ and $I_2( \omega_1, \omega_2,  \omega_3)$
defined in  Eqs. (\ref{GLE26}) and  (\ref{GLE27}) contribute. 
To leading order we obtain
  
\begin{align}
\label{Pole5}
I_1( \omega_1, \omega_2,  \omega_3) & \equiv 
\frac{ \imath  \left(\tau - \imath/\omega_1\right)^2 }
{2  \omega_1 (\omega_1-\omega_2)^2  (\omega_1-\omega_3)^2} \nonumber \\
&\simeq  \frac{ \imath  \left(\tau - \imath/\omega_1\right)^2 }
{2  \omega_1 \omega_2^2  \omega_3^2} \simeq  \frac{\gamma^2 m^2 \tau^2}{2 K^3(\gamma + K\tau)},
\end{align}
\begin{align}
\label{Pole6}
I_2( \omega_1, \omega_2,  \omega_3) 
&\equiv -
\frac{ 2\imath \left(\tau - \imath/\omega_2\right)  \left(\tau - \imath/\omega_3\right) }
{   (\omega_2 + \omega_3) (\omega_2-\omega_1)  (\omega_3-\omega_1) (\omega_2-\omega_3)^2}  \nonumber \\
&\simeq  -
\frac{ 2\imath \tau^2  }
{   (\omega_2 + \omega_3) \omega_2  \omega_3 (\omega_2-\omega_3)^2}   \nonumber \\
&\simeq   \frac{ m^2 \tau^5}{2 \gamma(\gamma + K\tau)}.
\end{align}
Inserting these results into  Eq. (\ref{GLE25}) we obtain 
\begin{align}
\label{Pole7}
\tau_{\rm rel} & \simeq   \frac{2K^2}{m^2 \tau^2} \left[ 
I_1( \omega_1, \omega_2,  \omega_3) +I_2( \omega_1, \omega_2,  \omega_3)  \right] \nonumber \\
& \simeq   \frac{2K^2}{m^2 \tau^2} \left[ 
 \frac{\gamma^2 m^2 \tau^2}{2 K^3(\gamma + K\tau)} +  \frac{ m^2 \tau^5}{2 \gamma(\gamma + K\tau)}  \right] \nonumber \\
 &= \frac{\gamma^2 }{ K(\gamma + K\tau)} +  \frac{ K^2 \tau^3}{ \gamma(\gamma + K\tau)}  \nonumber \\
 &=  \frac{\gamma}{K} \left( \frac{1 + (\tau K/\gamma)^3}{1+\tau K/\gamma}\right)
\end{align}
which is Eq. (\ref{GLE28}).

\section{Grote-Hynes Theory} \label{AppGH}

The Grote-Hynes (GH) theory  predicts the mean-first passage time to reach over the barrier,
as illustrated  in Fig. \ref{fig1}c),
in the presence of memory friction as \cite{Grote_1980}
\begin{equation}
\tau^{GH}_{{\rm mfp}} =\tau^{TST}   \frac{ \omega_{\text{max}}}{\lambda}
=   \frac{2 \pi \omega_{\text{max}}}{\lambda \omega_{\text{min}}}e^{\beta U_0},
\label{AppGH1}
\end{equation}
where $\omega_{\text{max}}= \sqrt{K_{\rm bar}/m}$ and 
$\omega_{\text{min}} = \sqrt{K/m}$ are the frequencies at the potential barrier and minimum
(with both curvatures defined as being positive) 
and $\tau^{TST}$ denotes  the   transition state theory  prediction.
$\lambda$ is the barrier reactive frequency, which is determined by  the GH equation
\begin{equation}
\lambda^2 + \lambda \frac{\hat{\Gamma}(\lambda)}{m} = \omega^2_{\text{max}},
\label{Eq:GHEquation}
\end{equation}
where $\hat{\Gamma}(\lambda)$ is the Laplace transform of the friction memory kernel. 
For a single exponential memory kernel given by Eq.~\eqref{GLE11} we obtain
\begin{equation}
\hat{\Gamma}(\lambda) = \int_0^\infty \Gamma(t)e^{-\lambda t} dt = \frac{\gamma}{1 +\tau \lambda}.
\label{Eq:LaplaceGamma}
\end{equation}
Inserting  \eqref{Eq:LaplaceGamma} into the GH equation \eqref{Eq:GHEquation}, we obtain a cubic polynomial.
In  the Markov limit $\tau\rightarrow 0$  we have $\hat{\Gamma}(\lambda)=\gamma$ and
 the solution of Eq.~\eqref{Eq:GHEquation} reads
\begin{equation}
\lambda = -\frac{\gamma}{2 m} \pm \left( \frac{\gamma^2}{4m^2} +\omega^2_{\text{max}}\right)^{1/2}.
\end{equation}
Taking the positive root, we obtain
\begin{equation} \label{GH7}
 \tau^{GH}_{{\rm mfp}} = \omega_{\text{max}}\left[\left(
\frac{\gamma^2}{4m^2} +\omega^2_{\text{max}}\right)^{1/2}-\frac{\gamma}{2 m}\right]^{-1} \tau^{TST},
\end{equation}
which is equivalent to the Kramers intermediate-to-high friction result.
 In the Markovian high-friction limit $\frac{\gamma}{m}\gg 1$ we obtain from Eq. (\ref{GH7})
\begin{equation}
 \tau^{GH}_{{\rm mfp}}  = \frac{2 \pi\gamma}{K |K_{\rm bar}|}e^{\beta U_0},
\end{equation}
which is identical to our result in Eq. (\ref{pot3}) provided we choose for the relaxation time the result
obtained in the Markovian mass-less limit $\tau_{rel} = \gamma/K$.
In the Markovian low-friction limit  $\frac{\gamma}{m}\ll 1$ we obtain from Eq. (\ref{GH7})
\begin{equation}
 \tau^{GH}_{{\rm mfp}}  = \tau^{TST}, 
\end{equation}
which means that in this limit GH theory becomes identical to transition state theory.
As explained in Sec. \ref{2pole}  the transition state theory prediction corresponds to the prediction 
of the Kramers theory at the crossover between the high-friction and low-friction, 
the Kramers turnover,
where the barrier crossing time has a minimum. Thus, the transtion-state-theory prediction 
is a lower bound for the barrier-crossing time, friction and energy-diffusion effects  increase the 
actual barrier-crossing time. 

In the zero-mass  $m\rightarrow 0$ limit
the solution of the GH Eq. \eqref{Eq:GHEquation} reads 
\begin{equation}
 \lambda = \frac{|K_{\rm bar}|/\gamma}{1 - \tau  |K_{\rm bar}|/\gamma},
\end{equation}
so that we obtain
\begin{equation}
\tau^{GH}_{\text{mfp}}=\frac{2\pi  \gamma (1 - \tau K_{\rm bar}/\gamma)}{K}  
\left(\frac{K}{K_{\rm bar}}\right)^{1/2}  e^{\beta U_0},
\end{equation}
which agrees for small memory time $\tau$ exactly with our result in Eq. \eqref{GLE29},
except that the barrier curvature $K_{\rm bar}$ (which enters the GH theory) has to be replaced
by $K$,  the curvature at the potential bottom. 
So the GH theory correctly recovers the speed-up regime for small  $\tau$ 
but misses the long-memory slow-down effect.

\section{ Derivation of  time-reversal symmetry of the Mori GLE \label{AppMori}}

In deriving a GLE, it is convenient (though not necessary) 
to consider a general  time-independent  Hamiltonian for  a system of $N$ interacting particles or atoms  in three-dimensional space of the form 
\begin{align}
\label{eq_Hamiltonian}
H (\omega) &= \sum_{n=1}^N \frac{\ve{p}_n^2}{2 m_n} + V ( \{ \ve r_N\}),
\end{align}
where a point in the 6$N$-dimensional  phase space  is denoted by 
$\omega = ( \{ \ve r_N\},  \{ \ve p_N\}  )$, which is a $6N$-dimensional  vector consisting
of the Cartesian particle positions  $\{ \ve r_N\}$ 
 and the  conjugate momenta  $ \{ \ve p_N\}$ and fully specifies the microstate of the system.
The Hamiltonian  splits into a kinetic and a potential part and  $m_n$ is the mass of particle  $n$. 
The potential $V ( \{ \ve r_N\})$ contains all interactions between the particles and 
includes possible external potentials. 
Using  the Liouville operator
\begin{align}
\label{eq_Liouville}
{\cal L}(\omega)&= \sum_{n=1}^N  \sum_{\alpha=x,y,z}  
\left(   \frac{ \partial H(\omega) }{\partial p_n^\alpha }  \frac{ \partial  }{\partial r_n^\alpha }  -
 \frac{ \partial H(\omega) }{\partial r_n^\alpha }  \frac{ \partial  }{\partial p_n^\alpha } 
 \right),
\end{align}
the Heisenberg observable $B(w, t)$ is defined by propagation according to \cite{Netz2025c}
\begin{align}
\label{eq_Heisenfilter} 
B(w, t)  \equiv  e^{(t-t_0){\cal L}(w)}  B_{S}(w),
\end{align}
 where $ B_{S}(w)$ is the ordinary time-independent observable, 
here denoted as Schrödinger observable in order to distinguish it from the Heisenberg observable.
At  time $t=t_0$ the Heisenberg observable coincides with the Schrödinger observable. 
In the following we assume that the observable $ B_{S}(w)$ only depends on particle positions, i.e. 
$ B_{S}(w) = B_S({\bf \{ r\}}_N)$.
The  time derivatives appearing in Eq. \eqref{rev1} follow from the propagator expression of $B(w, t)$ 
in Eq.~\eqref{eq_Heisenfilter} as
$ \dot B(w, t)= {\cal L}(w) B(w, t)$ and $ \ddot B(w, t)= {\cal L}^2(w) B(w, t)$. 
From the symmetry $H ( \{ \ve r_N\},  \{ \ve p_N\}  )= H ( \{ \ve r_N\},  -\{ \ve p_N\}  )$
of the Hamiltonian Eq.~\eqref{eq_Hamiltonian}, the  symmetry 
${\cal L}  ( \{ \ve r_N\},  \{ \ve p_N\}  ) =  - {\cal L}  ( \{ \ve r_N\}, - \{ \ve p_N\}  )$ of the 
Liouville operator Eq.~\eqref{eq_Liouville} 
and the propagator expression Eq.~\eqref{eq_Heisenfilter}, the time reversibility relations in Eq.~\eqref{rev2}
immediately follow.

The derivation of the GLE in Eq. \eqref{rev1}   follows standard procedures \cite{zwanzig_nonequilibrium_2001}:
One introduces a projection operator ${\cal P}$ that acts on phase space functions
 and its complementary projection operator ${\cal Q}$ via the relation $1= {\cal Q} + {\cal P}$.
 Inserting this unit operator into the time propagator relation Eq.~\eqref{eq_Heisenfilter},
 one obtains  after a few steps the GLE in Eq. \eqref{rev1},
where $t_P$ defines the time at which the projection is performed, which in principle can 
differ from the time $t_0$ at which the time propagation of the  Heisenberg 
variable in Eq.~\eqref{eq_Heisenfilter} starts \cite{Netz2025c}.

The  parameter $K_M$ in Eq. \eqref{rev1}  
corresponds to the potential stiffness  divided by the effective mass and 
 is given by 
\begin{align}
K_M &= \frac{\langle ( {\cal L}B(\omega, t_P) )^2 \rangle}
{\langle (B (\omega, t_P) - \langle B \rangle )^2 \rangle}. 
\end{align}
Here we define the expectation value of an arbitrary phase-space function $X(\omega) $ 
  with respect to a time-independent projection  distribution 
$ \rho_{\rm P}(\omega)$ as
\begin{align} \label{projectdist}
\langle   X(\omega)  \rangle = 
\int {\rm d}\omega   X (\omega) \rho_{\rm P}(\omega),
  \end{align}
  which we here take to be the normalized equilibrium canonical distribution
    $\rho_{\rm P}(\omega)= e^{-H(\omega)/(k_BT)}/ Z$, where  $Z$ is the  canonical partition function
    $Z=\int {\rm d}\omega e^{-H(\omega)/(k_BT)}$.
Obviously, the parameter $K_M$ is independent of time and particle momenta and therefore
not influenced by momentum or time reversal.

 The complementary  force in Eq. \eqref{rev1}  is given by
\begin{align}
\label{eq_F_operator}
F_M(\omega,t)&  \equiv  e^{(t-t_P) {\cal Q} {\cal L}} {\cal Q}{\cal L}^2  B(\omega, t_P)
\end{align}
and the memory friction kernel is given by \cite{Netz2025c}
\begin{align} 
\Gamma_M(t) = 
\frac{\langle F_M(\omega,0) F_M(\omega, t) \rangle}{\langle ( {\cal L}B(\omega, t_P) )^2 \rangle}.
 \label{eq_mori_memory}
\end{align}
From the fact that the Liouville operator is anti-self-adjoint and that the complementary projection operator ${\cal Q}$ 
is self-adjoint and idempotent, i.e. ${\cal Q}^2={\cal Q} $, it follows that 
the memory friction kernel  is  a symmetric function, i.e.   $\Gamma_M(t) = \Gamma_M(-t)$,
which is used to derive the time reversibility of the friction-memory term in Eq.~\eqref{rev3}. 

\bibliography{MemRelax}

\end{document}